\documentclass[%
preprint
amsmath,amssymb,
aps,
pra,
]{revtex4-1}

\usepackage{graphicx}% Include figure files
\usepackage{dcolumn}% Align table columns on decimal point
\usepackage{bm}% bold math
\usepackage{longtable}
\usepackage{ulem}

\usepackage{xcolor,soul}
\usepackage{mathptmx}
\usepackage{color,amsmath,amssymb,amsfonts}%
\usepackage{ccaption}

\newcommand*\Laplace{\mathop{}\!\mathbin\bigtriangleup}

\begin{document}

%\preprint{APS/123-QED}

\title{Rheology and Velocity Distribution of Couette Flow  for Protoplasmic Streaming }% 
% Force line breaks with \\

\author{Tomoya Souma  }
\email{tomoya@argopilot.com}
\affiliation{%
	Argopilot Coorporation, Tomiya, Japan% Force line breaks with \\ if necessary
}%
\author{Fumitake Kato}%
\email{katobnbf@ibaraki-ct.ac.jp}
\affiliation{
	National Institute of Technology (KOSEN), Ibaraki College, Hitachinaka, Japan% Force line breaks with \\ if necessary
}
\author{Hiroshi Koibuchi} % Write as First name Surname
\email[Corresponding author: ]{koibuchih@gmail.com}
%\affiliation{
	% Argopilot Coorporation, Tomiya, Japan% Force line breaks with \\ if necessary
	%}
\affiliation{
	National Institute of Technology (KOSEN), Ibaraki College, Hitachinaka, Japan,\\
	and
	 Argopilot Coorporation, Tomiya, Japan% Force line breaks with \\ if necessary
}

\begin{abstract}
We investigate two-dimensional Couette flow between parallel plates as a model of protoplasmic streaming in plant cells using a combined Langevin and Navier–Stokes simulation. The interaction between the fluid and biological materials is implemented using a Finsler geometry (FG) modeling technique, in which the interaction strength, represented by quantities such as relative viscosity, effectively varies with an internal degree of freedom (IDOF) associated with the biological materials. The simulations demonstrate that the Couette flow exhibits a linear velocity profile along the vertical direction perpendicular to the plates, as well as a peak structure in the velocity distribution consistent with reported experimental observations obtained by laser light scattering. The simulations also demonstrate that the model exhibits shear-thinning behavior typical of non-Newtonian fluids, with the viscosity minimized in the boundary region. Additionally, we find that the rheological properties are connected to the conformal transformation of the interaction between the fluid and the IDOF. Within the FG modeling framework, the conformally flat configuration of the IDOF, which is neither anisotropic nor randomly isotropic, is introduced as a minimum-energy state that characterizes the shear-thinning rheological properties of two-dimensional Couette flow.
\end{abstract}

\maketitle
% Section %%%%%%%%%%%%%%%%%%%%%%%%%%%%%%%%%%%%%%%%%%%%%%%%%%%%%%%%%%%%%%%%%%%%%%%%%
\section{Introduction\label{intro}}
\begin{figure}[ht]
	\centering{}\includegraphics[width=11.0cm,clip]{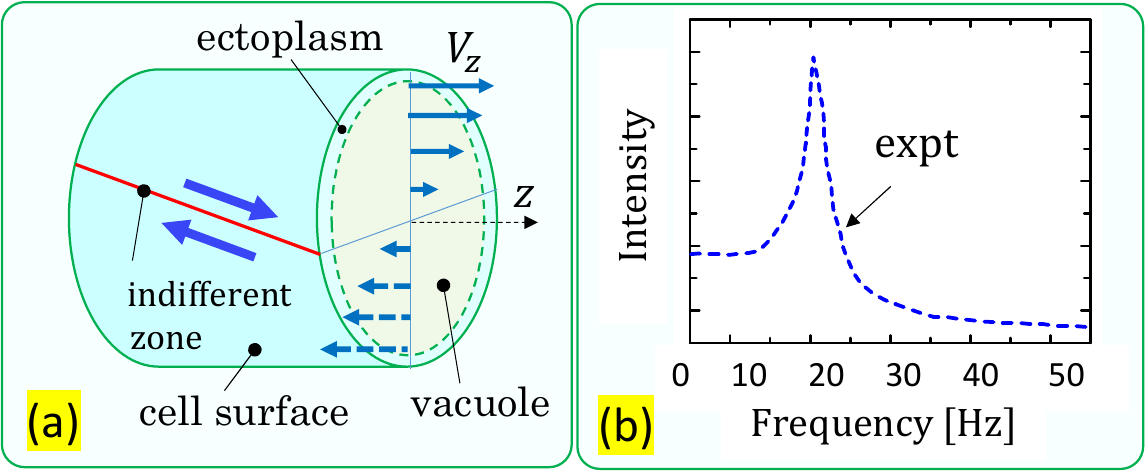}
	\caption{
		(a) An illustration of protoplasmic streaming, where the indifferent zone is indicated by a red line. (b) Experimentally observed distribution of velocity $h(V_z)$ as a function of Doppler shift frequency, which is proportional to $|V_z|$ \cite{Mustacich-Ware-PRL1974,Mustacich-Ware-BJ1976,Mustacich-Ware-BJ1977,Sattelle-Buchan-JCS1976}. The peak position indicates the speed of the molecular motors at the boundary.
		\label{fig-1}}
\end{figure}

%%%%%%%%%%%%%%%%%%%%%%%%%%%%%%%%%%%%%%%%%%%%%%%%%%%%%%%%%%%%%%%%%%%%%%%%%%%%%%%%%%%
Protoplasmic streaming is illustrated in Fig. \ref{fig-1}(a), where the flow direction is indicated by arrows on the cylindrical surface and along the vertical line in the cross section \cite{ShimmenYokota-COCB2004,VLubics-Goldstein-Protoplasma2009,TominagaIto-COPB2015}. The red line denotes an indifferent zone rotating on the cell surface.  The velocity distribution, such as $V_z$ in Fig. \ref{fig-1}(a), along a diameter of a cross section of Nitella cells was first measured by Kamiya and Kuroda using an optical microscope \cite{Kamiya-Kuroda-1956,Kamiya-Kuroda-1958,Kamiya-Kuroda-1973,Kamiya-1986,Tazawa-pp1968}. They found that the velocity in the ectoplasm (Fig. \ref{fig-1}(a)) is constant, while its magnitude non-linearly decreases toward the interior. 
The dashed line in Fig. \ref{fig-1}(b) represents an experimental result obtained by laser light scattering for the velocity distribution \cite{Mustacich-Ware-PRL1974,Mustacich-Ware-BJ1976,Mustacich-Ware-BJ1977,Sattelle-Buchan-JCS1976}. The vertical axis, labeled "intensity", indicates the number of particles moving with velocity $V_z$, corresponding to the "frequency" on the horizontal axis. The intensity peak indicates the maximum number of particles moving at the velocity corresponding to the peak position. This peak velocity is understood to represent the velocity of biological materials driven by molecular motors moving along actin filaments at the cell periphery \cite{Squires-Quake-RMP2005,McintoshOstap-CSatAG2016,Astumian-Science2020,Julicher-etal-RMP1997}. The velocity tail extending beyond the peak position toward higher velocities is attributed to the Brownian motion of particles, including molecular motors \cite{Ermark-McCammon-JCP1978,Brady-McCammon-ARFM1988,Roux-PhysA1992,Hossain-etal-POF2022}.
%%%%%%%%%%%%%%%% f-1
%%%%%%%%%%%%%%%

The flow velocity in plant cells attracts attention from the perspective of agricultural applications because it is closely related to plant size \cite{TominagaIto-COPB2015}. Therefore, the physical properties of cell components, such as the cytoskeleton and cell walls, as well as cytoplasmic streaming, have been studied as  biophysical research targets \cite{Amari-etal-BioPhysi2025,Goldstein-etal-JFM2010,Raymond-Goldstein-IF2015,Kikuchi-Mochizuki-PlosOne2015,Niwayama-etal-PNAS2010,Ganguly-Goldstein-etal-PNAS2012}. The effects of the wall-slip condition and cell elongation on the cytoplasmic streaming have been extensively studied \cite{Wolff-etal-Interface2012,Htet-Lauga-PNASNexus2025}.

Specifically, the flow profile should also be characterized in detail from a fluid-mechanical perspective. Goldstein et al. investigated  flows governed by the Stokes equations coupled with an advection-diffusion equation for a scalar field representing a metabolite, with particular focus on the role of circulating flows in the indifferent zone in enhancing mixing \cite{Goldstein-etal-PRL2008,Goldstein-etal-PNAS2008}.

The rheological property is expected to be non-Newtonian or remains almost Newtonian, as discussed in \cite{Kikuchi-Mochizuki-PlosOne2015,Niwayama-etal-PNAS2010,Ganguly-Goldstein-etal-PNAS2012}. Many biological materials, such as polymeric materials \cite{doi2013soft,Doi-Edwards-1986}, are present in cytoplasmic streaming. It is well known that polymer solutions are typical examples of non-Newtonian fluids \cite{Schroeder-JRheo2018,Chabra-Richardson-2025,Dinic-etal-Wiley2022}. In cytoplasmic streaming in Drosophila oocytes, frequency-dependent viscosity has been measured, and the results indicate non-Newtonian behavior at short timescales \cite{Ganguly-Goldstein-etal-PNAS2012}. Since the activation force is generated by molecular motors that share a common mechanism with those in animal cells, shear-thinning behavior, in which the viscosity decreases with increasing shear rate, is also expected in plant cells.

Egorov et al. and Noro et al. numerically studied protoplasmic streaming using two-dimensional (2D) Couette flow between parallel plates \cite{Egorov-etal-POF2020} and three-dimensional (3D) flows in a cylindrical domain with spiraling and non-spiraling boundary conditions \cite{Noro-etal-PHYSA2023}, respectively. They used Langevin Navier–Stokes (LNS) simulations to examine whether the velocity distribution shown in Fig. \ref{fig-1}(b) could be reproduced. In the LNS equations, a Gaussian random force is included as an external force to model fluid fluctuations, following the Landau–Lifshitz framework.  In Ref. \cite{Noro-etal-PHYSA2023}, it was shown that the 3D spiraling flow reproduces the experimentally observed velocity distribution when an appropriate strength of the random force is applied.

However, the LNS simulations in \cite{Egorov-etal-POF2020,Noro-etal-PHYSA2023} did not incorporate any internal degrees of freedom (IDOFs) associated with biological materials, such as metabolites, and the fluid was assumed to be Newtonian. Consequently, the rheological properties of the fluid were not considered, although its non-Newtonian nature cannot be completely disregarded, as mentioned above \cite{Ganguly-Goldstein-etal-PNAS2012}.

In this paper, we investigate the rheological properties of protoplasmic streaming by introducing a directional degree of freedom, $\vec{\tau} \in S^1/2$ (${\rm half\text{-}unit\;circle}$), as an internal degree of freedom (IDOF) for biological materials within the framework of Finsler geometric (FG) modeling \cite{Koibuchi-Sekino-PhysA2014,Proutorov-etal-JPC2018,FKato-etal-PhysA2026,SS-Chern-AMS1996}. The coupling between the IDOF and the flow field is implemented using a hybrid simulation that combines Monte Carlo (MC) updates of $\vec{\tau}$ with the LNS simulations. The elastic properties arising from cytoskeletal networks are not taken into account. 

The purpose of this study is to demonstrate the shear-thinning property of the streaming, which is closely connected with enhanced mixing. Additionally, we demonstrate that the linear dependence of $V_x$ on $y$ in 2D Couette flow is not influenced by the IDOF and that the normalized velocity distribution of $V_x$ is similar to the experimentally observed distribution reported in laser light scattering studies.

% Section %%%%%%%%%%%%%%%%%%%%%%%%%%%%%%%%%%%%%%%%%%%%%%%%%%%%%%%%%%%%%%%%%%%%%%%%%
\section{Computations of Couette flow\label{2D-Couette}}
%%%%%%%%%%%%%%%%%%%%%%%%%%%%%%%%%%%%%%%%%%%%%%%%%%%%%%%%%%%%%%%%%%%%%%%%%%%%%%%%%%%
% Subsection %%%%%%%%%%%%%%%%%%%%%%%%%%%%%%%%%%%%%%%%%%%%
\subsection{Lattice structure\label{lattice-structure}}
%%%%%%%%%%%%%%%%%%%%%%%%%%%%%%%%%%%%%%%%%%%%%%%%%%%%%%%%%
%%%%%%%%%%%%%%%% f-2
\begin{figure}[h!]
	\centering{}\includegraphics[width=11.0cm,clip]{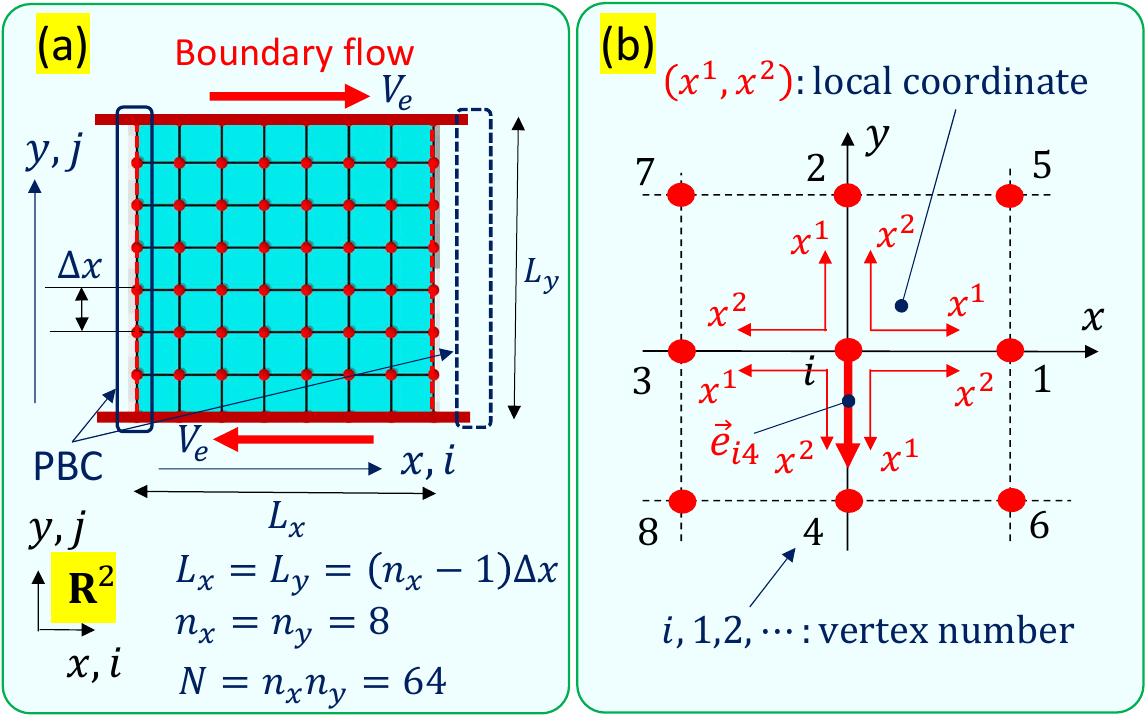}
	\caption{
(a) Computational domain on a regular square lattice. The small size, $N\!=\!n_x\times n_y\!=\!64$, is used to show the lattice structure. (b) Four possible local coordinate systems $(x_1,x_2)$ at vertex $i$. The symbol $i$ in (a) denotes the position along the $x$ axis, while $i$ in (b) denotes the vertex number.
		\label{fig-2}}
\end{figure}
%%%%%%%%%%%%%%%
For the Couette flow, a 2D regular square lattice is used, as shown in Fig. \ref{fig-2}(a). Periodic boundary conditions are imposed in the horizontal, or $x$, direction for the flow field and the IDOF. The IDOF is assumed to be located at the lattice sites, implying that it moves with the velocities of both the internal streaming and the boundary flows. The lattice geometry used in the simulations is given by
\begin{eqnarray}
	\label{lattice-geometry}
	n_x=n_y=40,\quad N=1600, \quad{\rm (regular\; square \; lattice)}.
\end{eqnarray}
The lattice spacing $\varDelta x$ and the system size, $L_x\!=\!(n_x\!-\!1)\varDelta x\!=\!L_y$, are determined by the diameter of a plant cell, as described below. The position of each vertex is denoted by $(x,y)$ or $(i,j)$, representing the Cartesian coordinates. The symbol $i$ is also used to denote the vertex index, as shown in Fig. \ref{fig-2}(b). Four possible orientations of the local coordinate axes $(x_1,x_2)$ at vertex $i$ are shown together with the Cartesian coordinates $(x,y)$. As described below, these local coordinate systems are used to define the Finsler metric. Specifically, the unit vectors $\vec{e}_{ij}, (j\!=\!1,\!\cdots\!,4)$  along the local coordinate axes are used to calculate the unit Finsler length $\chi_{ij}$ (Appendix \ref{App-A}).

% Subsection %%%%%%%%%%%%%%%%%%%%%%%%%%%%%%%%%%%%%%%%%%%%
\subsection{Review of 3D lattice results \label{3D-lattice-results}}
\begin{figure}[h!]
	\centering{}\includegraphics[width=11.0cm,clip]{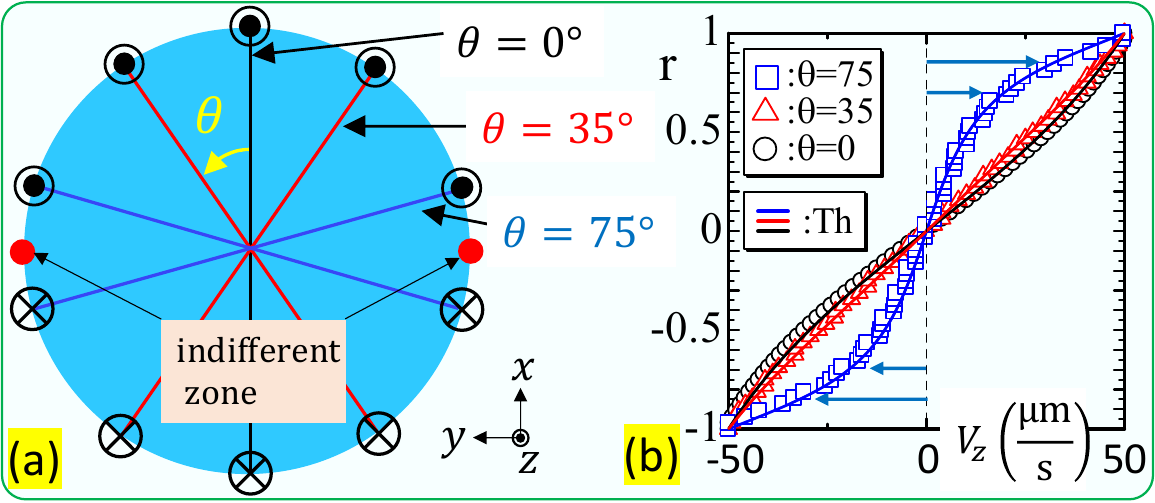}
	\caption{
		(a) Cross section of a 3D tube, with the indifferent zone indicated by red circles ($\textcolor{red}{\bullet}$). The straight lines indicate the $\theta$-dependent positions at which $V_z$ is calculated. (b) The simulated and theoretical results for the $\theta$ dependence of the radial variation of $V_z$ reported in Refs. \cite{Noro-etal-PHYSA2023,Pickard-CJB1971}.
		\label{fig-3}}
\end{figure}

%%%%%%%%%%%%%%%%%%%%%%%%%%%%%%%%%%%%%%%%%%%%%%%%%%%%%%%%%
Figure \ref{fig-3}(a) shows a cross section of the three-dimensional cylindrical domain used in the simulations reported in Ref. \cite{Noro-etal-PHYSA2023}. The indifferent zone is indicated by the red symbols ($\textcolor{red}{\bullet}$) at the boundary. The flow is inward (outward) in the upper (lower) half of the domain, as indicated by $\odot$ ($\otimes$). The lines at $\theta\!=\!0^\circ$, $35^\circ$, and $75^\circ$ indicate the locations where the velocity component $V_z$ is evaluated. The resulting velocity profiles are shown in Fig. \ref{fig-3}(b), together with the theoretical predictions \cite{Pickard-CJB1971}. From the $V_z$ profiles shown in Fig. \ref{fig-3}(b), we find that the dependence of $V_z$ on $r$ differs significantly from that in Couette flow, in which $V_x$ is assumed to depend linearly on $y$, even for non-Newtonian fluids. The nonlinear dependence of $V_z$ on $r$ arises from three-dimensional effects in the flow field, suggesting that two-dimensional Couette flow is not suitable for capturing such effects.
%%%%%%%%%%%%%%%% f-3
%%%%%%%%%%%%%%%

However, experimentally prepared flows between two parallel plates or within two cylindrical tubes cannot always be regarded as 2D flows in the case of non-Newtonian fluids containing polymeric materials with 3D degrees of freedom \cite{note-1}. Such non-Newtonian fluids are expected to be influenced by the three-dimensional effects of the constituent materials. Therefore, it is reasonable to investigate whether the incorporation of an internal degree of freedom through Finsler geometry can induce a shear thinning even in 2D Couette flow, using the 2D Langevin Navier–Stokes equations.

% Subsection %%%%%%%%%%%%%%%%%%%%%%%%%%%%%%%%%%%%%%%%%%%%
\subsection{Plant cell parameters in physical and simulation units\label{plant-cell}}
%%%%%%%%%%%%%%%%%%%%%%%%%%%%%%%%%%%%%%%%%%%%%%%%%%%%%%%%%
Physical quantities characterizing the protoplasmic streaming are the fluid density $\rho_e ({\rm kg/m^3})$, kinematic viscosity $\nu_e ({\rm m^2/s})$, boundary velocity $V_e ({\rm m/s})$, and diameter $d_e ({\rm m})$, where $d_e$ corresponds to the separation distance between the parallel plates of the Couette flow. 
 These quantities are listed in Table \ref{table-1}. 
It should be noted that
\begin{eqnarray}
	\label{rho-and-nu}
	\rho_e=\rho_{\rm water},\quad 
       \nu_e=\left\{ \begin{array}{@{\,}ll}
		10 \,\nu_{\rm water}\quad  \;\; ({\rm case\;A}) \\
		100 \,\nu_{\rm water}\quad  ({\rm case \;B}) 
	\end{array} \right.
\end{eqnarray}
are assumed. In case A, $\nu_e$ is 10 times smaller than $\nu_e$ in case B, whose value is identical to that assumed in Refs. \cite{Egorov-etal-POF2020,Noro-etal-PHYSA2023}.

Simulation units can be introduced using positive scaling parameters $\alpha$, $\beta$, and $\gamma$ for length, time, and mass, respectively, such that $V_e [{\rm m/s}]\!=\!V_0 [{\rm \alpha m/(\beta s)}]$, $\nu_e [{\rm m^2/s}]\!=\!\nu_0 [{\rm (\alpha m)^2/(\beta s)}]$, and $\rho_e [{\rm kg/m^3}]\!=\!\rho_0 [{\rm \gamma kg/(\alpha m)^3}]$, where $V_0\!=\!V_e \beta/\alpha$, $\nu_0\!=\!\nu_e \beta/\alpha^2$, and $\rho_0\!=\!\rho_e \alpha^3/\gamma$. The parameters $V_0$, $\nu_0$, and $\rho_0$ are also listed in Table \ref{table-1}.
%++++++++++++++++++++++++++++++++++
\begin{table}[h!]
	\caption{
		Physical parameters $\nu_{e}$, $V_{e}$, and $d_{e}$, $\cdots$, characterize the protoplasmic streaming in plant cells. The symbols with the subscript $e$ denote experimentally observed or observable quantities expressed in physical units, whereas those with the subscript $0$ are expressed in simulation units.
		\label{table-1}} 
	\begin{center}
		\begin{tabular}{ccccccccccccc}
			\hline
			\hline
			case && $\rho_e ({\rm kg/m^3})$ && $\nu_{e}~({\rm m^2}/{\rm s})$ && $V_{e}~({\rm m}/{\rm s})$ && $d_{e}~({\rm m})$ &&  $\varDelta x~({\rm m})$ &&  $\varDelta t~({\rm s})$    \\
			\hline
			\hline
			A && $1\times 10^{3}$  && $1\times 10^{-5}(=\!10 \,\nu_{\rm water})$  && $5\!\times\!10^{-5}$ && $5\!\times\!10^{-4}$ && $1.25\times\!10^{-5}$  && $5\!\times\!10^{-4}$  \\
			B && $1\times 10^{3}$  && $1\times 10^{-4}(=\!100 \,\nu_{\rm water})$  && $5\!\times\!10^{-5}$ && $5\!\times\!10^{-4}$ && $1.25\times\!10^{-5}$  && $5\!\times\!10^{-5}$  \\
			\hline
			\hline
			case && $\rho_0~[\frac{\rm \gamma kg}{\rm (\alpha m)^3}]$ && $\nu_0~[\frac{\rm (\alpha m)^2}{\rm \beta s}]$ &&\ $V_0~[\frac{\rm \alpha m}{\rm \beta s}]$ && $d_{0}~(\alpha {\rm  m})$ && $\varDelta x_0~[{\rm \alpha m}]$ && $\varDelta t_0~[{\rm \beta s}]$ \\
			\hline
			A && $1\!\times\! 10^{-3}$  &&  $1\times 10^{6}$  && $5$ && $500$ &&  $12.5$ && $5\times10^{-6}$   \\
			B && $1\!\times\! 10^{-3}$  &&  $1\times 10^{7}$  && $5$ && $500$ &&  $12.5$ && $5\times10^{-7}$   \\
			\hline
			\hline
		\end{tabular}
	\end{center}
\end{table}
%++++++++++++++++++++++++++++++++++

The parameters $\alpha$, $\beta$, and $\gamma$ are expressed in terms of the ratios between the physical and simulation units of the relevant quantities as
\begin{eqnarray}
	\label{numbers-for-units}
	\alpha=\frac{\nu_e}{\nu_0}\frac{V_0}{V_e}, \quad \beta=\frac{\nu_e}{\nu_0}\left(\frac{V_0}{V_e}\right)^2, \quad \gamma=\frac{\rho_e}{\rho_0}\left(\frac{\nu_e}{\nu_0}\frac{V_0}{V_e}\right)^3,
\end{eqnarray}
and are listed in Table \ref{table-2}. The parameter $D$, which characterizes the Brownian random force, is fixed at $D\!=\!10$ (case A) and $D\!=\!100$ (case B) in the simulations. These values are determined by the fluctuation-dissipation relation, which is discussed in Appendix \ref{App-B}. The Reynolds number is given by
\begin{eqnarray}
	\label{Reynolds-number}
R_e=\frac{V_e d_e}{\nu_e}=\left\{ \begin{array}{@{\,}ll}
	2.5\times 10^{-3}\quad  ({\rm case\;A}) \\
	2.5\times 10^{-4}\quad  ({\rm case \;B}) 
\end{array} \right..
\end{eqnarray}

%++++++++++++++++++++++++++++++++++
\begin{table}[htb]
	\caption{
		Parameters $\alpha$, $\beta$, $\gamma$, and $D$ assumed in the simulations, expressed in simulation units, and the Reynolds number. $D_{\rm fd}\!=\!9.8$ for case A and $D_{\rm fd}\!=\!98$ for case B are expected from the fluctuation-dissipation relation (Appendix \ref{App-B}) and are close to $D\!=\!10$ and $D\!=\!100$, respectively.
		 \label{table-2} }
	\begin{center}
		\begin{tabular}{ccccccccccccc }
			\hline
			\hline
			case && $\alpha$ && $\beta$ && $\gamma$ && $D ({\rm {[\alpha m]^2}{[\beta s]^{-3}}})$ &&  $D_{\rm fd}({\rm {[\alpha m]^2}{[\beta s]^{-3}}})$ && $R_e$ \\
			\hline
			A &&  $1\times 10^{-6}$ &&  $0.1$  &&  $1\!\times\!10^{-12}$ && 10  && 9.8 && $2.5\!\times\!10^{-3}$ \\
			B&&  $1\times 10^{-6}$ &&  $0.1$  &&  $1\!\times\!10^{-12}$ && 100  && 98 && $2.5\!\times\!10^{-4}$ \\
			\hline
			\hline
		\end{tabular}
	\end{center}
\end{table}
%++++++++++++++++++++++++++++++++++

% Section %%%%%%%%%%%%%%%%%%%%%%%%%%%%%%%%%%%%%%%%%%%%%%%%%%%%%%%%%%%%%%%%%%%%%%%%%
\section{Simulation technique \label{sim-tech}}
%%%%%%%%%%%%%%%%%%%%%%%%%%%%%%%%%%%%%%%%%%%%%%%%%%%%%%%%%%%%%%%%%%%%%%%%%%%%%%%%%%%
% Subsection %%%%%%%%%%%%%%%%%%%%%%%%%%%%%%%%%%%%%%%%%%%%
\subsection{Langevin Navier-Stokes equation and SMAC method \label{LNS-eq}}
%%%%%%%%%%%%%%%%%%%%%%%%%%%%%%%%%%%%%%%%%%%%%%%%%%%%%%%%%
The LNS equation is given by
\begin{eqnarray}
	\label{LNS-eq}
	\begin{split}
		\frac{\partial \vec{V}}{\partial t} &=-\left(\vec{V}\cdot \nabla(\tau)\right)\vec{V}-\rho^{-1}\nabla p+\nu \Laplace (\tau) \vec{V}+ \vec{\eta},\\
		\nabla\cdot\vec{V}&=0.
	\end{split}
\end{eqnarray}
The final term $\vec{\eta}\left(=\!(\eta^1,\eta^2)\right)$ on the right-hand side represents a Gaussian random force satisfying
\begin{eqnarray}
	\label{Gaussian-random-force}
	\langle \eta^a(\vec{r},t)\eta^b(\vec{r}^{\,\prime},t^\prime)\rangle = 2D\delta^{a b}\delta(\vec{r}-\vec{r}^{\,\prime})\delta(t-t^\prime),
\end{eqnarray}
where $D$ is a positive parameter corresponding to those listed in Table \ref{table-2}. The symbol $\tau$ in $\nabla(\tau)$ and $\Laplace (\tau)$ indicates that the differential operators are modified by the IDOF within the FG modeling procedure (Appendices \ref{App-A} and \ref{App-C}). The advection term $-\left(\vec{V}\cdot \nabla(\tau)\right)\vec{V}$ on the right-hand side of Eq. (\ref{LNS-eq}) can be neglected because the Reynolds numbers in Eq. (\ref{Reynolds-number}) are very small. However, we retain this term for completeness (Appendix \ref{App-D}).

To ensure $\nabla\cdot\vec{V}\!=\!0$, we use the SMAC method as follows \cite{McKee-etal-CandF2007}. The discrete forms of Eq. (\ref{LNS-eq}) are obtained by introducing an intermediate velocity $\vec{V}_i^*$ such that
\begin{eqnarray}
	\label{LNS-temporal-1}
	&&{\vec V}_i^*(t)= \vec{V}_i(t) +\varDelta t \left[ \left (-{\vec V}_i\cdot \nabla(\tau)\right){\vec V}_i(t)-{\rho}^{-1} {\it \nabla} p_i(t) +\nu \Laplace(\tau) {\vec V}_i(t)\right] +\sqrt{2D\varDelta t} \,\vec{g}_i(t),\\
	\label{LNS-temporal-2}
	&&{\vec V}_i(t+\varDelta t)={\vec V}_i^*(t)-\varDelta t {\rho}^{-1} {\it \nabla} \left[ p_i(t+\varDelta t)-p_i(t)\right].
\end{eqnarray}
Taking $\nabla\cdot$ of Eq. (\ref{LNS-temporal-2}), we find that
\begin{eqnarray}
	\label{divless-cond}
\nabla \cdot \vec{V}_i(t+\varDelta t)=0, 
\end{eqnarray}
provided that 
\begin{eqnarray}
	\label{Smac-Poiss}
	\Laplace \phi_i(t)=\frac{\rho}{\varDelta t}\nabla\cdot{\vec V}_i^*(t),\quad \phi_i(t) = p_i(t+\varDelta t)-p_i(t)
\end{eqnarray}
are satisfied. The first is a Laplace equation, which is solved using an iterative technique with an acceleration coefficient of 1.88 for the lattice of size in Eq. (\ref{lattice-geometry}), and the second is the definition of the scalar function $\phi_i$.  Note that $\nabla$ acting on $p$ in Eqs. (\ref{LNS-eq}),(\ref{LNS-temporal-2}), and  $\nabla$ in Eq. (\ref{divless-cond}), and $\Laplace$ in Eq. (\ref{Smac-Poiss}) are standard operators that are independent of $\vec{\tau}$.

The discrete form of Eq. (\ref{Gaussian-random-force}) is given by
\begin{eqnarray}
	\label{disc-GRF}
		\langle \eta_i^a(t)\eta_j^b(t^\prime)\rangle = 2D\delta^{a b}\delta_{ij}\delta(t-t^\prime),
\end{eqnarray}
where the subscript $i$ denotes a lattice site. The symbol $\vec{g}_i(t)\!=\!(g_i^1,g_i^2)$ in Eq. (\ref{LNS-temporal-1}) denotes Gaussian random numbers with mean 0 and variance 1, satisfying $\langle g_i^a\rangle\!=\!0$ and $\langle (g_i^{a})^2\rangle\!=\!1$ for $(a\!=\!1,2)$.
The relation between $\vec{g}_i(t)$ and $\vec{\eta}_i(t)$ is given by \cite{Egorov-etal-POF2020}
\begin{eqnarray}
	\label{Gaussian-random-eta-and-g}
	\vec{\eta}_i(t)=\sqrt{\frac{2D}{\varDelta t}}\vec{g}_i(t).
\end{eqnarray}
It is reasonable to consider that the random force $\eta_i^a(t)$ corresponds to the divergence of the fluctuating stress tensor $s_{ab}$ introduced by Landau and Lifshitz \cite{Landau-Lifschitz-StatPhys}, such that
\begin{eqnarray}
	\label{eta-div-of-sab}
	\eta^a_i(t)=\rho^{-1}\frac{\partial s_{ab}}{\partial x_b}.
\end{eqnarray}
Therefore, using the fluctuation-dissipation relation  for $s_{ab}$, as mentioned above, we obtain a relation between the parameter $D$ and the kinematic viscosity $\nu$.  From the discussion in Appendix \ref{App-B}, the values  $D_{\rm fd}\!=\!9.8$ and  $D_{\rm fd}\!=\!98$ are expected, and therefore, we assume
\begin{eqnarray}
	\label{assumed-D}
	D=10 \; ({\rm case\; A}\Leftrightarrow \nu_e\!=\!10 \,\nu_{\rm water}),\quad D=100 \; ({\rm case\; B}\Leftrightarrow \nu_e\!=\!\!100 \,\nu_{\rm water})
\end{eqnarray}
in the simulations as shown in Table \ref{table-2}. It should be noted that our LNS simulation technique can be compared with that based on dissipative particle dynamics  \cite{Hoogerbrugge-Koleman-EPL1992,Espanol-Warren-EPL1995,Espanol-EPL1997,Groot-Warren-JCP1997,Hoogerbrugge-Koelman-EPL1992,Nikunen-etal-CPC2003}.

% Subsection %%%%%%%%%%%%%%%%%%%%%%%%%%%%%%%%%%%%%%%%%%%%
\subsection{Hamiltonian for Monte Carlo update of IDOF\label{Hamiltonian-for-MC}}
%%%%%%%%%%%%%%%%%%%%%%%%%%%%%%%%%%%%%%%%%%%%%%%%%%%%%%%%%
%%%%%%%%%%%%%%%% f-4
\begin{figure}[h!]
	\centering{}\includegraphics[width=8.0cm,clip]{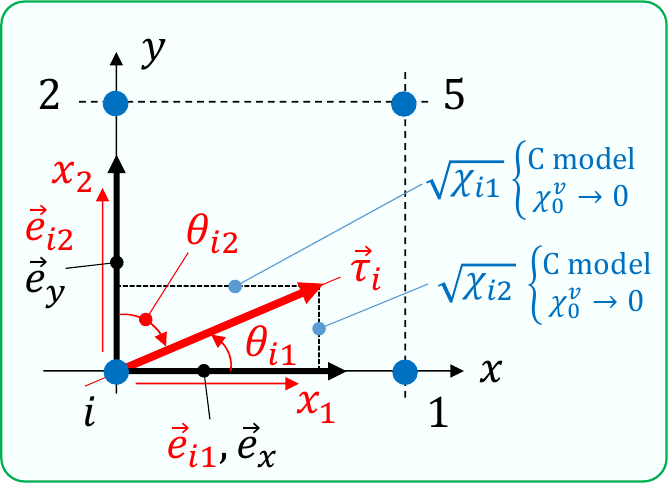}
	\caption{
		IDOF variable $\vec{\tau}_i$ at vertex $i$, where $(x_1,x_2)$ denote local coordinate axes which are identical with the Cartesian coordinate axes $(x,y)$.  The symbols $\vec{e}_{i1}(=\!\vec{e}_{x})$ and $\vec{e}_{i2}(=\!\vec{e}_{y})$  denote unit vectors along the $x_1$ and $x_2$ axes, respectively. The symbol $\theta_{i1}$ ($\theta_{i2}$)  denotes the angle between the $x_1$ ($x_2$) axis and $\vec{\tau}_i$.
		\label{fig-4}}
\end{figure}
%%%%%%%%%%%%%%%
The viscous energy $H_V$, which describes an energy dissipation in fluid, is given by (Appendix \ref{App-A})
\begin{eqnarray}
	\label{discrete-HV-main}
	\begin{split}
	H_V&= \sum_{ij}\Gamma_{ij}(\tau)(\vec{V}_i-\vec{V}_j)^2, \\
	\Gamma_{i1}&=\bar{\Gamma}^{-1}\left(\frac{\chi_{i1}}{\chi_{i2}}+\frac{\chi_{1i}}{\chi_{15}}\right), \quad  \Gamma_{i2}=\bar{\Gamma}^{-1}\left(\frac{\chi_{i2}}{\chi_{i1}}+\frac{\chi_{2i}}{\chi_{25}}\right),
\end{split}
\end{eqnarray}
where the unit Finsler lengths $\chi_{i1}$ and $\chi_{i2}$ can be defined in two different models; C model and S model,  such that  
\begin{eqnarray}
	\label{discrete-chi-for-HV-main}
	\begin{split}
		\chi_{i1}&=\left\{ \begin{array}{@{\,}ll}
			\quad|\vec{\tau}_i\cdot\vec{e}_{i1}|^2+\chi_0^v\;\;\;=\cos^2\theta_{i1}+\chi_0^v\quad  ({\rm C\; model})  \\
			1-(\vec{\tau}_i\cdot\vec{e}_{i1})^2+\chi_0^v=\sin^2\theta_{i1}+\chi_0^v\quad  ({\rm S\;model})
		\end{array} \right.,\\
		\chi_{i2}&=\left\{ \begin{array}{@{\,}ll}
			\quad|\vec{\tau}_i\cdot\vec{e}_{i2}|^2+\chi_0^v\;\;\;=\cos^2\theta_{i2}+\chi_0^v\quad  ({\rm C\; model}) \\
			1-(\vec{\tau}_i\cdot\vec{e}_{i2})^2+\chi_0^v=\sin^2\theta_{i2}+\chi_0^v\quad  ({\rm S\;model})
		\end{array} \right., \\ \vec{\tau}_i&\in S^1/2\; ({\rm half\text{-}unit\;circle}), \quad \chi_0^v=0.5.
	\end{split}
\end{eqnarray}
The IDOF $\vec{\tau}_i$, the angles  $\theta_{i1}$, $\theta_{i2}$ and unit vectors $\vec{e}_{i1}$, $\vec{e}_{i2}$ are illustrated in Fig. \ref{fig-4}. As mentioned in the Introduction, $\vec{\tau}_i$ represents a directional degree of freedom of biological material at site $i$. 
The system of 2D Couette flow is in its non-equilibrium steady state, and therefore, the mean value of this energy $H_V$  remains constant.

An interaction between the flow field and the biological materials is implemented in the intensive part $\Gamma_{ij}(\tau)$ using a Finsler metric. The detailed information on $\Gamma_{ij}(\tau)$ in $H_V$ is given in Appendix \ref{App-A}.  The relative viscosity $\Gamma_{ij}({\tau})$ depends on $\vec{\tau}$, and therefore, the viscous energy $H_V$ with the coefficient $\xi$ can be included in the the total Hamiltonian $H(\tau)$ for the update of  $\vec{\tau}$.  The coefficient $\xi$ in this case is not always identical to $\nu_0$, because $\nu_0$ corresponds to the shear force exerted by neighboring fluid elements to drive the fluid, whereas $\xi$ corresponds to a shear force acting on materials in the flow field.  However, the coefficient $\Gamma_{ij}(\tau)$ is used in the Laplace operator associated with the viscous term in the LNS equation given in Eq. (\ref{LNS-eq}), as demonstrated in Eq. (\ref{discrete-Laplace}). For this reason, we refer to $\Gamma_{ij}(\tau)$ as the relative viscosity.  
Additionally,  since biological materials are present in protoplasmic streaming, the energy supplied in the fluid by the boundary flow is expected to be partly reflected in the kinetic energy $H_K$ of the materials. For the kinetic energy of non-point-like materials, rotational energy can also be included. However, the rotational motion of materials in flow fields is too complex to analyze. Therefore, we simply introduce the IDOF $\vec{\tau}$ as a directional degree of freedom of the materials.

Thus,  we assume that the Hamiltonian $H(\tau)$ for MC update of $\vec{\tau}$ is given by
\begin{eqnarray}
	\label{total-Hamiltonian}
	\begin{split}
	H(\tau)&= H_K+\xi H_V,\quad (\xi=0,\; {\rm or}\;\xi=0.5),
	\\
	H_K&=\frac{1}{2}m_0\sum_i m_{i}(\tau)\vec{V}_i\cdot \vec{V}_i, \quad
			m_{i}(\tau)=\overline {m}^{-1}\frac{1}{\chi_{i1}\chi_{i2}}, \quad m_0=0.004,
	 \end{split}
\end{eqnarray}
where
\begin{eqnarray}
	\label{discrete-chi-for-HK-main}
	\begin{split}
		\chi_{i1}&=\left\{ \begin{array}{@{\,}ll}
			\quad|\vec{\tau}_i\cdot\vec{e}_{i1}|^2+\chi_0^m\;\;\;=\cos^2\theta_{i1}+\chi_0^m\quad  ({\rm C\; model})  \\
			1-(\vec{\tau}_i\cdot\vec{e}_{i1})^2+\chi_0^m=\sin^2\theta_{i1}+\chi_0^m\quad  ({\rm S\;model})
		\end{array} \right.,\\
		\chi_{i2}&=\left\{ \begin{array}{@{\,}ll}
			\quad|\vec{\tau}_i\cdot\vec{e}_{i2}|^2+\chi_0^m\;\;\;=\cos^2\theta_{i2}+\chi_0^m\quad  ({\rm C\; model}) \\
			1-(\vec{\tau}_i\cdot\vec{e}_{i2})^2+\chi_0^m=\sin^2\theta_{i2}+\chi_0^m\quad  ({\rm S\;model})
		\end{array} \right., \\ \chi_0^m&=0.01.
	\end{split}
\end{eqnarray}
Note that $\chi_{i1}$ and $\chi_{i2}$ in Eq. (\ref{discrete-chi-for-HK-main}) for $H_K$ are the same as those in Eq. (\ref{discrete-chi-for-HV-main}) for $H_V$ except for the difference between $\chi_0^{m}$ and $\chi_0^{v}$. The interaction between the flow field and materials is geometrically implemented in $m_i(\tau)$ as a relative mass using the same technique as that used for $\Gamma_{ij}(\tau)$ in $H_V$. The detailed information of this $m_i(\tau)$ is given in Appendix \ref{App-C}.

We should note that $\vec{\tau}$ is the variable updated using $H(\tau)$, while the velocity $\vec{V}$ is fixed at the fluid velocity. Therefore, in the MC update of $\vec{\tau}$, the boundary velocity is assumed to be the same as that in the LNS simulations such that 
\begin{eqnarray}
	\label{V-normalization}
V_0({\rm MC})= V_0({\rm LNS})=5.
\end{eqnarray}
This relation does not mean that the units defined  by $\alpha$, $\beta$ and $\gamma$ in Eq. (\ref{numbers-for-units}) for the LNS simulation are applicable to the MC unit.

The simulation unit for this MC update process is $k_BT\!=\!1$,  where $k_BT$ corresponds to an energy scale of $4\times 10^{-21}\,({\rm J})$ at room temperature. The parameter $m_0$ in $H_K$ is fixed at $m_0\!=\!1/250\!=\!0.004$, and therefore, we have a kinetic energy  of the material such that
\begin{eqnarray}
	\label{HK-energy-V0}
	 \frac{H_K}{N}=\frac{1}{2} m_0V_0^2\times k_BT=0.05\, k_BT.
\end{eqnarray}
Here, we assume that the biological material is a cube with side length $\varDelta \ell$ and mass density $\rho(=10^3 [\frac{\rm kg}{\rm m^3}])$, which is the same as that of water. Therefore, for materials moving with $V\!=\!V_e$, we find from the relation
\begin{eqnarray}
	\label{HK-energy}
    \frac{1}{2}mV^2=\frac{1}{2}\rho {(\varDelta \ell)^3 V_e^2}=\frac{1}{2} m_0V_0^2\times k_BT\left(=0.05\, k_BT\right),
\end{eqnarray}
that  $(\varDelta \ell)^3\!=\!\frac{m_0 V_0^2 \times k_BT}{\rho V_e^2}\!=\!\frac{m_0 V_0^2 \times k_BT}{\rho V_e^2}\!=\!\frac{m_0\times k_BT}{\rho}(\frac{V_0}{V_e})^2  \!=\!\frac{0.004\times 4\times 10^{-21}}{10^3}\times 10^{10}\!=\!1.6\times 10^{-16}$, and $\varDelta \ell\!=\!(0.16\times 10^{-15})^{1/3}\!\simeq\! 0.54\times 10^{-5}({\rm m})\!=\!5.4 ({\rm \mu m})$. 
The size $5.4 ({\rm \mu m})$ is considered reasonable for biological materials in Chara plant cells. 
 When the boundary velocity is varied to $V_e\!>\!50 ({\rm \frac{\mu m}{s}})$ or $V_e\!<\!50 ({\rm \frac{\mu m}{s}})$ while keeping ${V_0}/{V_e}(=\!10^{5})$ fixed, the corresponding size of materials moving with $V\!=\!V_e$ remains unchanged because the size depends only on $m_0$.

The energy scale of $\xi H_V, (\xi\!=\!0.5)$ is comparable to that of $H_K$, so its inclusion in $H(\tau)$ is reasonable. The continuous form of $H_V$ is given by $H_V\!=\!\int d^2x (\frac{\partial \vec{V}}{\partial x}\cdot\frac{\partial \vec{V}}{\partial x}\!+\!\frac{\partial \vec{V}}{\partial y}\cdot\frac{\partial \vec{V}}{\partial y})$, which can be written as $H_V\!=\!\int d^2x (\frac{\partial V_x}{\partial y})^2$, because $\vec{V}\!=\!(V_x,0)$ and $\partial V_x/\partial  x\!=\!0$ for 2D Couette flow between parallel plates. Therefore, the energy per bond, $\frac{H_V}{2N}$, is estimated as $(\frac{dV_x}{dy})^2\!=\![V_0/(n_x/2)]^2\!=\![5/(n_x/2)]^2\!=\!(5/20)^2\!=\!0.0625 (k_BT)$, based on the assumption that the lattice spacing is fixed at 1 in the MC update of $\vec{\tau}$, in contrast to $\varDelta x$ in the LNS iteration for $\vec{V}$. Thus, we obtain 
\begin{eqnarray}
	\label{Hv-energy}
	\frac{\xi H_V}{2N}\simeq 0.031\,k_BT,\quad(\xi=0.5),
\end{eqnarray}
which is comparable to $H_K/N$ in Eq. (\ref{HK-energy}).

% Subsection %%%%%%%%%%%%%%%%%%%%%%%%%%%%%%%%%%%%%%%%%%%%
\subsection{Hybrid simulation technique \label{hybrid-simulation}}
%%%%%%%%%%%%%%%%%%%%%%%%%%%%%%%%%%%%%%%%%%%%%%%%%%%%%%%%%
The time evolution of a set of velocities $\{\vec{V}_i(t),\, (i={\rm inerior\; vertices)}\}$, denoted by $\{\vec{V}_i(t)\}$, is obtained by integrating the discrete LNS equations given by Eqs. (\ref{LNS-temporal-1}), (\ref{LNS-temporal-2}), and (\ref{Smac-Poiss}). However, due to the random force $\{\vec{g}_i\}$ and the coupling with the IDOF $\{\vec{\tau}_i\}$, the converged configuration of $\{\vec{V}_i\}$  still fluctuates depending on the initial values of $\{\vec{V}_i\}$ and $\{\vec{g}_i\}$. Therefore, averaging is necessary for the converged configurations. The variable  $\vec{\tau}_i$ is updated using the Metropolis MC technique such that $\vec{\tau}_i\to \vec{\tau}^\prime_i$ is accepted with the probability ${\rm Min}[1,\exp(-\delta H)]$, where $\delta H\!=\!H(\vec{\tau}^\prime_i)\!-\!H(\vec{\tau}_i)$ using the Hamiltonian $H$ described in Eq. (\ref{total-Hamiltonian})  \cite{Metropolis-JCP-1953,Landau-PRB1976}.  The procedure for averaging a physical quantity $Q$, which depends on $\{\vec{V}_i\}$ and  $\{\vec{\tau}_i\}$, is as follows:
\begin{enumerate}
	\item [$\mathrm{(i)}$] $\{\vec{V}_i^*(t)\}$ is calculated from the current $\{\vec{V}_i(t)\}$, $\{p_i(t)\}$ and $\{\vec{g}_i\}$ as in Eq. (\ref{LNS-temporal-1}). $\{\phi_i(t)\}$ is obtained by solving Poisson's equation in Eq. (\ref{Smac-Poiss}). The iteration technique is used to  solve the Poisson's equation with an acceleration coefficient 1.88 and the convergent criterion
\begin{eqnarray}
 	\label{convergent-criterion}
 	{\rm Max}\{|\phi_i({\rm new})-\phi_i({\rm old})|\}<1\times 10^{-10}.
 \end{eqnarray}
    Then, $\{\vec{V}_i(t+\varDelta t)\}$ and $\{p_i(t+\varDelta t)\}$ are calculated by Eq. (\ref{LNS-temporal-2}) and $\{p_i(t)\!+\!\phi_i(t)\}$, respectively.
	\item [$\mathrm{(ii)}$] MC update $\vec{\tau}_i\to \vec{\tau}^\prime_i$, ($i\!=\!1,\cdots,N$), is performed  with the probability ${\rm Min}[1,\exp(-\delta H_K)]$.
	\item [$\mathrm{(iii)}$] Steps $\mathrm{(i)}$ and $\mathrm{(ii)}$ are repeated $n_{\rm MC}$ times;
\begin{eqnarray}
	\label{nmc}
	n_{\rm MC}=1\times 10^5.
\end{eqnarray}
	\item [$\mathrm{(iv)}$] Step $\mathrm{(i)}$ is performed until the convergent criteria are satisfied:
\begin{eqnarray}
	\label{convergent-criteria}
	\begin{split}
	&{\rm Max}\{|\vec{V}_i(t+\varDelta t)-\vec{V}_i(t)|\}<1\times 10^{-8},\\
	&{\rm Max}\{|p_i(t+\varDelta t)-p_i(t)|\}<1\times 10^{-8}.
    \end{split}
\end{eqnarray}
	\item [$\mathrm{(v)}$] Steps $\mathrm{(i)}$ to $\mathrm{(iv)}$ are repeated $n_{\rm itr}$ times to obtain the mean value of physical quantity $Q$  from the converged $\{Q_i\}$;
\begin{eqnarray}
	\label{nitr}
	Q=\frac{1}{n_{\rm itr}} \sum_{i=1}^{n_{\rm itr}} Q_i, \quad (n_{\rm itr}=500).
\end{eqnarray}
\end{enumerate}

We should note that $\vec{\tau}$ is influenced by $\vec{V}$ through $H_K$ and $H_V$ in the MC update process $\mathrm{(ii)}$, while $\vec{V}$ is influenced by $\vec{\tau}$ through the viscous term $\Laplace (\tau) \vec{V}$ corresponding to $H_V$ in the time integration process $\mathrm{(i)}$ of the LNS equation. Therefore, the interaction between $\vec{\tau}$ and $\vec{V}$  through the kinetic energy $H_K$ is one-way, or non-reciprocal, in the sense that $\vec{V}$ in $H_K$ influences  $\vec{\tau}$,  solely in the MC update process, because the advection term $-\left(\vec{V}\cdot\nabla(\tau)\right)\vec{V}$, corresponding to $H_K$, is negligibly small during the LNS time-evolution process.  

% Section %%%%%%%%%%%%%%%%%%%%%%%%%%%%%%%%%%%%%%%%%%%%%%%%%%%%%%%%%%%%%%%%%%%%%%%%%
\section{Results\label{results}}
%%%%%%%%%%%%%%%%%%%%%%%%%%%%%%%%%%%%%%%%%%%%%%%%%%%%%%%%%%%%%%%%%%%%%%%%%%%%%%%%%%%
\subsection{Velocity distributions and rheology\label{V-distribution}}
%%%%%%%%%%%%%%%% f-5
\begin{figure}[h!]
	\centering{}\includegraphics[width=10.0cm,clip]{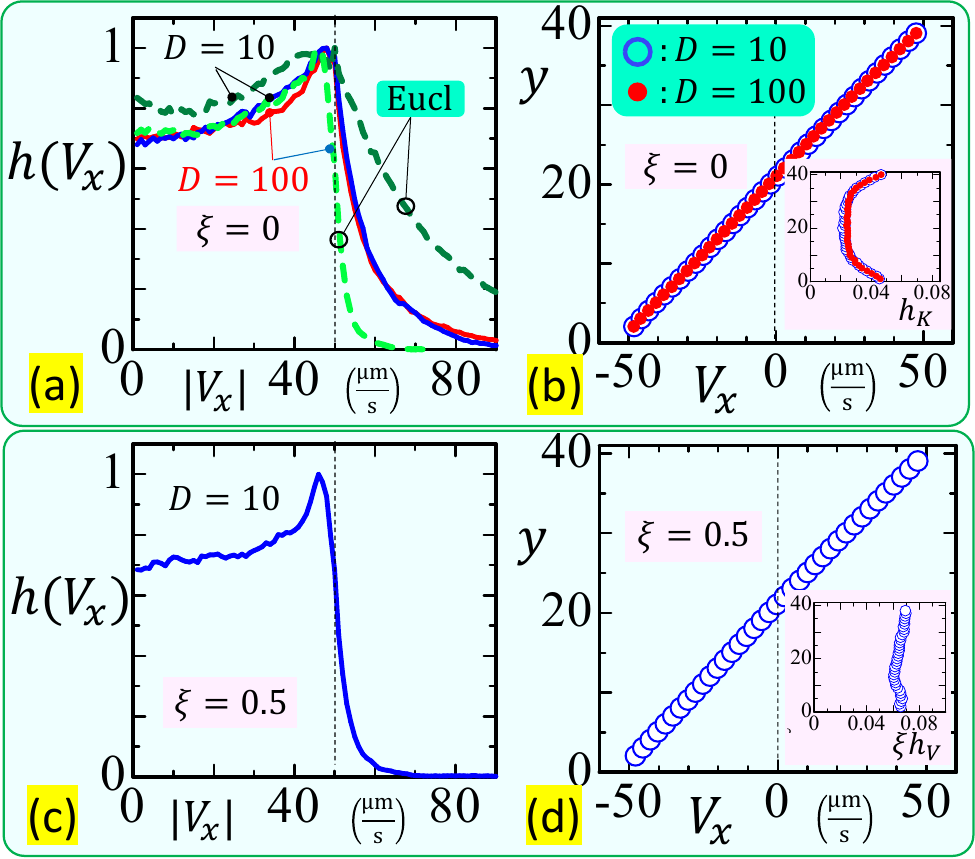}
	\caption{
		(a) Normalized histogram $h(V_x)$ of $|V_z|$ and (b) velocity distribution $V_x(y)$ versus $y$ obtained for $\xi\!=\!0$. (c) $h(V_x)$ versus $|V_z|$ and (d) $V_x(y)$ versus $y$ obtained for $\xi\!=\!0.5$. The dashed lines labeled by "Eucl" in (a) denote the results of the model without the IDOF $\vec{\tau}$.  $h_K/N$ and $\xi h_V/2N$ are shown inside the plots in (b) and (d), respectively. C model is used, and the boundary velocity is fixed at $V_e\!=\!50\,({\rm \mu m/s})$.
		\label{fig-5}}
\end{figure}
%%%%%%%%%%%%%%%
Using 500 MC samples obtained with C model, the histogram $h(V_x)$ of $|V_x|$;
\begin{eqnarray}
	\label{histogram-of-Vx}
	h(V_x)={\rm normalized \; distribution\; of}\; |V_x|,
\end{eqnarray}
and the $y$ dependence of $V_x(y)$ are calculated.  These are plotted in Figs. \ref{fig-5}(a) and \ref{fig-5}(b) for  $\xi\!=\!0$ with $D\!=\!10$ and $D\!=\!100$, and in Figs. \ref{fig-5}(c) and \ref{fig-5}(d) for $\xi\!=\!0.5$ with $D\!=\!10$. The boundary velocity is fixed at $V_e\!=\!50 ({\rm \mu m /s})$ as shown in Table \ref{table-1}. To evaluate the influence of the IDOF on $h(V_x)$, the simulation results obtained using the model without $\vec{\tau}$ are plotted as dashed lines in Fig. \ref{fig-5}(a) and labeled "Eucl". The shape of $h(V_x)$ for the Euclidean model is strongly influenced by variations in $D$, similar to the behavior observed in the 3D Euclidean flow models in Refs. \cite{Noro-etal-PHYSA2023}, in sharp contrast to the FG model, for which $h(V_x)$ is almost unaffected by variations in $D$. A linear dependence of $V_x$ on $y$ is also observed in the Euclidean model, as in the FG model shown in Fig. \ref{fig-5}(b).

The shape of $h(V_x)$ is similar to the result of the 3D simulations in Ref. \cite{Noro-etal-PHYSA2023} and different from those of the 2D simulations in Ref. \cite{Egorov-etal-POF2020}. Since the IDOF is not included in the model of Ref. \cite{Egorov-etal-POF2020}, this result suggests that the IDOF influences the shape of $h(V_x)$. 
Linear behavior, observed in $V_x(y)$ in Figs. \ref{fig-5}(b) and \ref{fig-5}(d), implies that the IDOF does not influence the linearity in $V_x(y)$ of the Couette flow  at least under the assumed simulation condition. Additionally, we find from the plots of $h_K(=\!H_K/N)$ and $\xi h_V(=\!\xi H_V/2N)$ as insets in Figs. \ref{fig-5}(b) and \ref{fig-5}(d) that the energy scale of $h_V$ is comparable to that of $h_K$ and  the expectation in Eq. (\ref{Hv-energy}).

%%%%%%%%%%%%%%%% f-6
\begin{figure}[h!]
	\centering{}\includegraphics[width=10.0cm,clip]{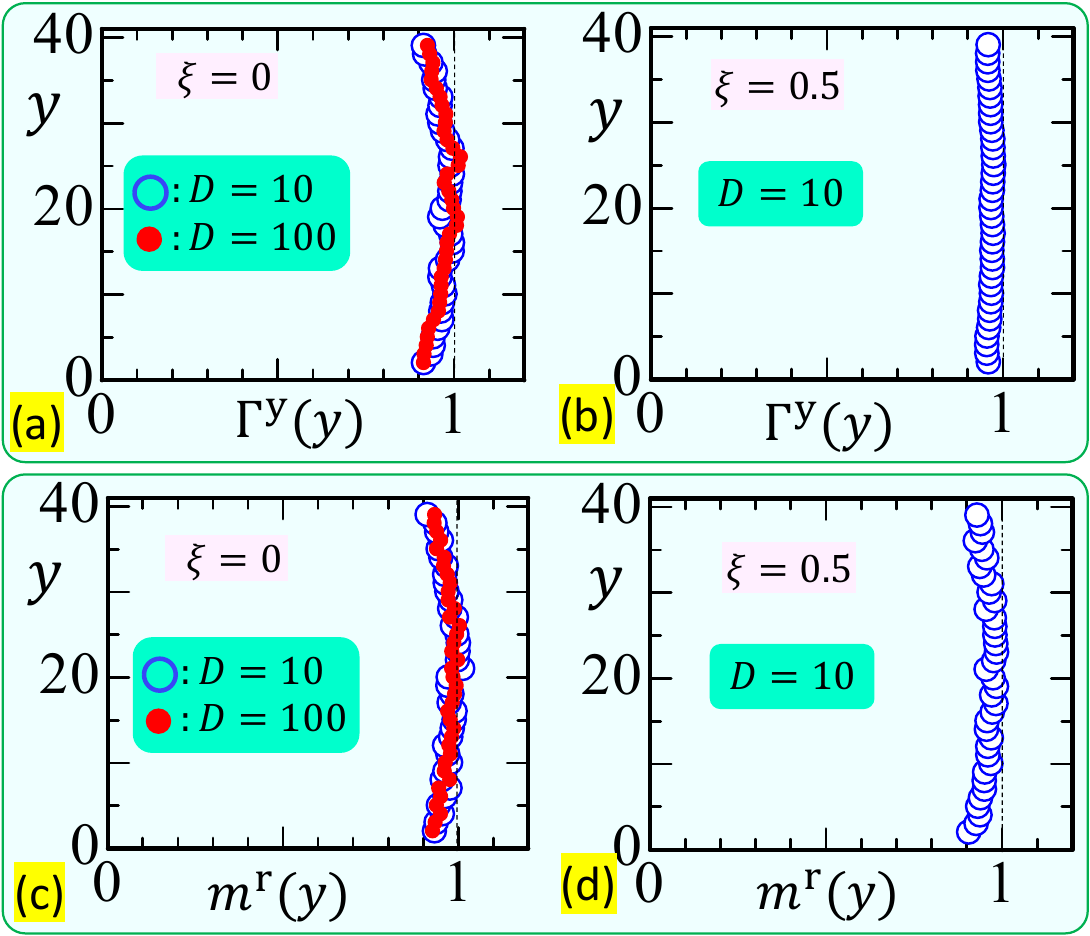}
	\caption{
		Relative viscosity $\Gamma_y(y)$ vs. $y$ for (a) $\xi\!=\!0$ with  $D\!=\!10$ (\textcolor{blue}{$\bigcirc$}) and $D\!=\!100$ (\textcolor{red}{$\bullet$}), and (b) $\xi\!=\!0.5$ with $D\!=\!10$.  Relative mass $m^r(y)$ vs. $y$  for (c) $\xi\!=\!0$ with  (\textcolor{blue}{$\bigcirc$}) and $D\!=\!100$ (\textcolor{red}{$\bullet$}), and (d) $\xi\!=\!0.5$ with $D\!=\!10$. The boundary velocity is fixed at  $V_e\!=\!50\,({\rm \mu m/s})$.
		\label{fig-6}}
\end{figure}
%%%%%%%%%%%%%%%
To characterize the rheological property, the relative viscosities $\Gamma_{x}(y)$, $\Gamma_{y}(y)$ are defined by
\begin{eqnarray}
	\label{relative-viscosity}
	\begin{split}
	\Gamma_{x}(y)&=\frac{1}{n_x}\sum_{i=1}^{n_x}\Gamma_{i1}(y=j),\quad
	\Gamma_{y}(y)=\frac{1}{n_x}\sum_{i=1}^{n_x}\Gamma_{i2}(y=j),
	\end{split}
	\end{eqnarray}
where $\sum_{i=1}^{n_x}$ denotes the summation over $i$ with fixed $j(=\!y)$. Here, the symbol $j$ denotes the lattice site $(i,j)$, and  $n_x(=\!\sum_{i1}1)$ is the total number of bond $i1$ along the $x$ axis, and $\Gamma_{i1}$ and $\Gamma_{i2}$ are given in Eq. (\ref{discrete-HV-main}) and Eq. (\ref{discrete-HV}). Note that $\Gamma_{x,y}(y)$ are defined only inside the boundaries ($\Leftrightarrow 1\!<\!y\!<\!n_x$). These quantities $\Gamma_{x}(y)$ and $\Gamma_{y}(y)$  are the intensive part of the viscous energy $H_V$ in Eqs. (\ref{discrete-HV-main}) and (\ref{discrete-HV}). We find from Figs. \ref{fig-6}(a),(b)  that $\Gamma_y(y)\!\leq\!1$ and $\Gamma_y(y)$ reaches its minimum at a position close to the boundary, and that this tendency is apparent in the case $\xi\!=\!0$ in Fig. \ref{fig-6}(a). We should note that $\Gamma_{x}(y)$, which is not plotted,  is similar to $\Gamma_{y}(y)$. The absence of a difference between  $\Gamma_{x}(y)$ and $\Gamma_{y}(y)$ indicates that there is no specific alignment or anisotropy in the IDOF $\vec{\tau}$. 

The relative mass $m^{\rm r}(y)$ at $y$ is defined by
\begin{eqnarray}
	\label{relative-mass}
	m^{\rm r}(y)=\frac{1}{n_x}\sum_{i=1}^{n_x}m_i (y),
\end{eqnarray}
where $m_i$ is given in Eqs. (\ref{total-Hamiltonian}) and (\ref{disc-kinetic-energy}), and the subscript $i$ denotes the site $(i,j)$ corresponding to the fixed position $j\!=\!y$. We find from Fig. \ref{fig-6}(c),(d) that $m^{\rm r}(y)\!\leq\!1$ over the entire range of $y$, and that the shape of $m^{\rm r}(y)$ is similar to that of $\Gamma_y(y)$. The result $m^{\rm r}(y)\!\leq\!1$ implies that the effective particle mass is reduced, especially near the boundary. We should note that $\Gamma_y(y)$ and $m^{\rm r}(y)$ are dimensionless quantities, and neither corresponds to a real physical quantity such as viscosity or mass.

%%%%%%%%%%%%%%%% f-7
\begin{figure}[h!]
	\centering{}\includegraphics[width=10.0cm,clip]{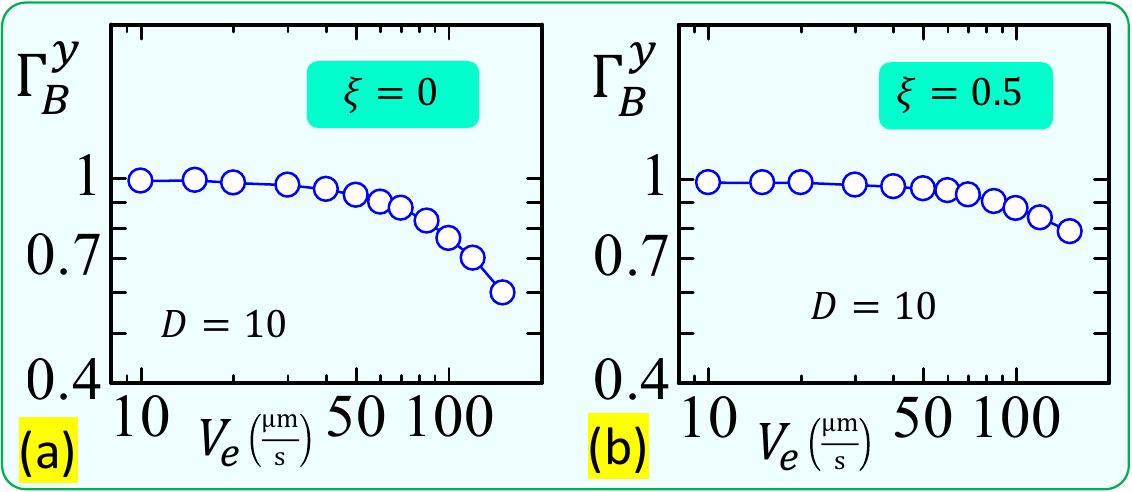}
	\caption{
		Relative viscosity $\Gamma_B^y$ vs. $V_e$ for (a) $\xi\!=\!0$ with $D\!=\!10$, (b) $\xi\!=\!0.5$ with $D\!=\!10$, both on a log-log scale. $V_e$ is proportional to the shear rate, because the height $L_y$ (Fig. \ref{fig-2}(a)) is fixed.
		\label{fig-7}}
\end{figure}
%%%%%%%%%%%%%%%
Finally, in this subsection, the relative viscosity $\Gamma^y_B$ versus $V_e$ is plotted in Figs. \ref{fig-7}(a),(b) on a log-log scale for $10\!\leq\!V_e\!\leq\!150$, where $\Gamma^y_B$ is the value of $\Gamma_y(y)$ at the boundary, defined by
\begin{eqnarray}
	\label{nuc-vs-Ve}
	\Gamma^y_B=\frac{1}{2}\left[\Gamma_y(j\!=\!3)+\Gamma_y(j\!=\!38)\right], \quad (38=n_y-2).
\end{eqnarray}
Note that $V_0$ is also varied while keeping $V_e/V_0$ fixed when $V_e$ is varied, so that $\alpha$, $\beta$, and $\gamma$ in Eq. (\ref{numbers-for-units}) remain unchanged. Varying the boundary velocity $V_e$ changes the shear rate because the cell diameter $d_e$, corresponding to the height $L_y$ (Fig. \ref{fig-2}(a)), is fixed. We find that $\Gamma^y_B$ decreases with increasing $V_e$, at least for $V_e\geq\! 50 ({\rm \mu m/s})$. 

\subsection{Snapshots and IDOF configuration \label{snapshots}}
%%%%%%%%%%%%%%%% f-8
\begin{figure}[h!]
	\centering{}\includegraphics[width=11.0cm,clip]{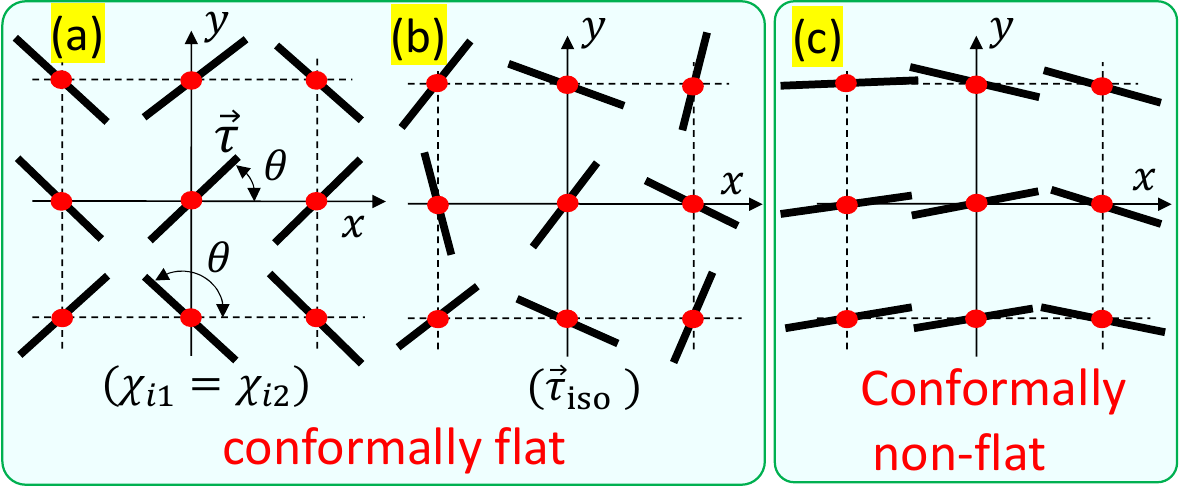}
	\caption{
(a) A conformally flat ($\Leftrightarrow$ $\theta\!=\! \pi/4$ or $\theta\!=\! 3\pi/4$) configuration of $\vec{\tau}$, a conformally flat (random isotropic) configuration, denoted by $\vec{\tau}_{\rm iso}$, and (c) a conformally non-flat (anisotropic) configuration of $\vec{\tau}$.
		\label{fig-8}}
\end{figure}
%%%%%%%%%%%%%%%
To examine the $\theta$ dependence of $m(\theta)$ in Eq. (\ref{relative-mass-theta}), we assume $\chi_0^m\!=\!0$. In this case, $m(\theta)$ is given by $m(\theta)\!=\!\bar{m}^{-1}\cos^{-2}\theta\sin^{-2}\theta$. It is then easy to check that $m(\theta)$ attains its minimum at $\theta\!=\!\pi/4$ and $\theta\!=\!3\pi/4$, for which the corresponding $\vec{\tau}$ configurations are depicted in Fig. \ref{fig-8}(a). Therefore, we obtain $\chi_{i1}\!=\!\chi_{i2}$, which is the condition for $g_{ij}$ in Eq. (\ref{F-metric-2D}) to be conformally flat, implying an isotropic, non-random interaction between the flow field and the biological materials. An example of an isotropic but random configuration is shown in Fig. \ref{fig-8}(b), where the mean value of $m_{i}(\tau)$ satisfies $m_{i}(\tau)\!\to\!1$ due to the normalization factor ${\bar m}^{-1}$. The difference between the configurations in  Figs. \ref{fig-8}(a) and \ref{fig-8}(b) is that $\chi_{i1}\!=\!\chi_{i2}$ holds locally at each site $i$ in Fig. \ref{fig-8}(a), whereas  $\chi_{i1}\!=\!\chi_{i2}$ holds only in terms of the mean values of $\chi_{ij}$ in Fig. \ref{fig-8}(b). As a consequence of this property, the kinetic energy $H_K$ for the conformally flat configuration  shown in Fig. \ref{fig-8}(a) is smaller than that for the randomly isotropic configuration shown in Fig. \ref{fig-8}(b). This will be confirmed below.  An example of a conformally non-flat ($\Leftrightarrow$ $\chi_{i1}\!\not=\! \chi_{i2}$) configuration corresponding to an anisotropic interaction is shown in Fig. \ref{fig-8}(c).

Here, we should note that the kinetic energy $H_K$ in Eq. (\ref{disc-kinetic-energy}) is minimized when $m_i$ is minimized because the extensive part $\vec{V}_i\cdot\vec{V}_i$ is primarily determined by the boundary velocity $\vec{V}_e$ independent of  $\vec{\tau}$ in 2D Couette flow. For this reason, it is expected that the $\vec{\tau}$ configuration satisfies $\theta\!\simeq\!\pi/4$ or $\theta\!\simeq\!3\pi/4$ in the boundary region, where $\vec{V}_i\cdot\vec{V}_i$ is relatively larger than in the central region even for non-zero $\chi_0^m\!=\!0.01$. It should also be emphasized that these configurations of $\vec{\tau}$ for the minimum-energy state of $H_K$ remain unchanged when $H_V$ is included with $\xi\!=\!0.5$ in the boundary region. The reason is that the extensive part $\vec{V}_i\cdot \vec{V}_i$ of $H_K$ is much larger than the extensive part $(\vec{V}_j\!-\!\vec{V}_i)^2$ of $H_V$ in that region.

\begin{figure}[h!]
	\centering{}\includegraphics[width=11.0cm,clip]{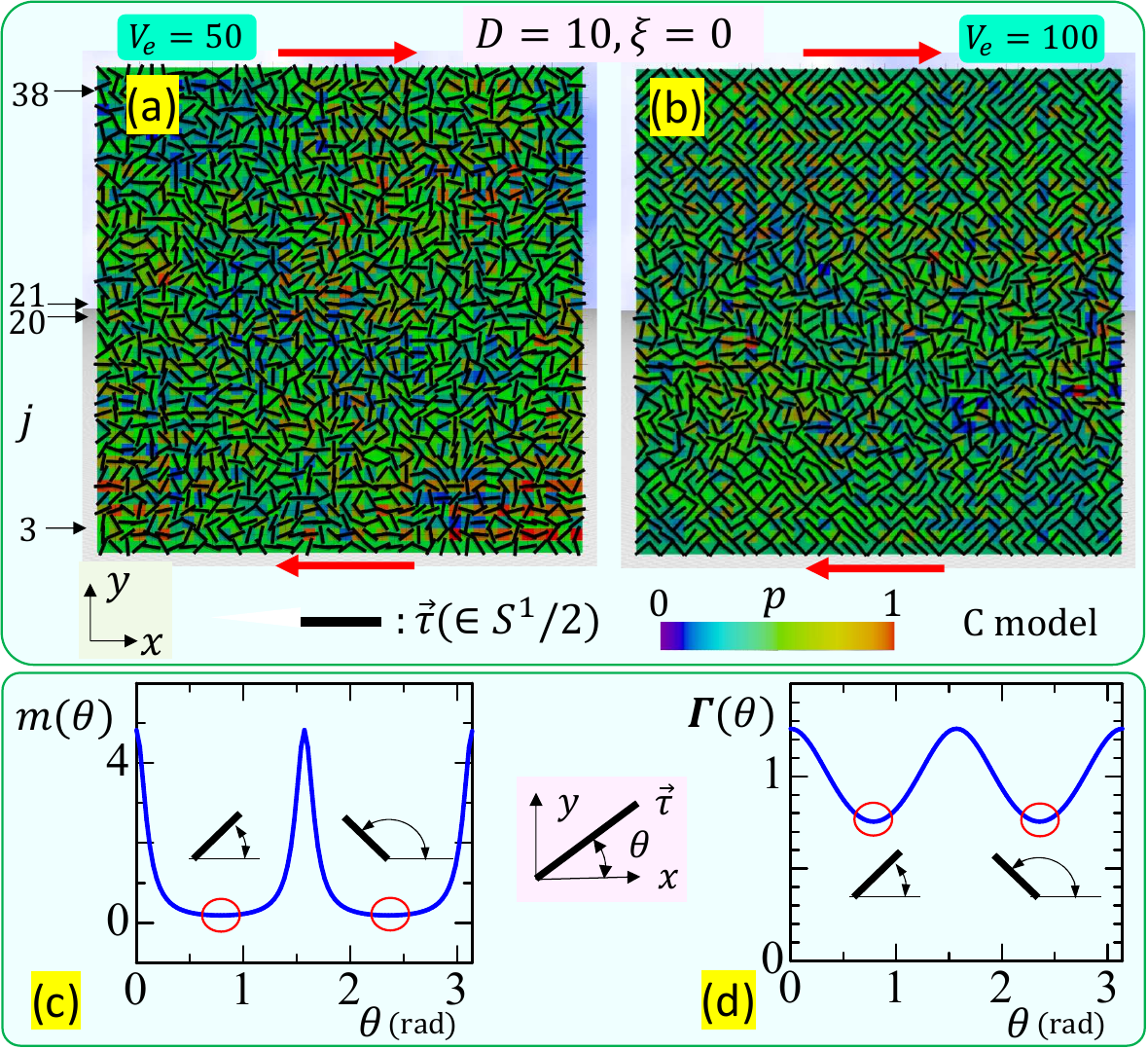}
	\caption{
		Snapshot of $\vec{V}$, the IDOF $\vec{\tau}$, and the normalized pressure $p$ obtained for (a)  $V_e\!=\!50\,({\rm \mu m/s})$, and  (b) $V_e\!=\!100\,({\rm \mu m/s})$ under $D\!=\!10$ and $\xi\!=\!0$. (c) Relative mass $m(\theta)$ defined in Eq. (\ref{relative-mass-theta}) versus $\theta$, and (d) relative viscosity ${\boldsymbol \varGamma}(\theta)$ defined in Eq. (\ref{relative-viscosity-TH}) versus $\theta$.
		\label{fig-9}}
\end{figure}

To check this expectation, a snapshot of a converged configuration IDOF $\vec{\tau}$, and normalized pressure $p$ for $V_e\!=\!50\,({\rm \mu m/s})$ is shown in Fig. \ref{fig-9}(a). We find that the alignment angle $\theta$ of $\vec{\tau}$ with respect to the $x$ axis is $\theta\!\simeq\!\pi/4$ or $\theta\!\simeq\!3\pi/4$ for many IDOF vectors. The alignment is not always uniform, and therefore, the configuration is not globally anisotropic. The configuration of $\vec{\tau}$ in Fig. \ref{fig-9}(b) for $V_e\!=\!100\,({\rm \mu m/s})$ is more prominent that $\theta\!\simeq\!\pi/4$ or $\theta\!\simeq\!3\pi/4$ is almost satisfied in the boundary region, while in the central region the configuration is comparable to that in Fig.  \ref{fig-9}(a) for $V_e\!=\!50\,({\rm \mu m/s})$.
%%%%%%%%%%%%%%%% f-9
%%%%%%%%%%%%%%%
The IDOF configuration for $\xi\!=\!0.5$ at $V_e\!=\!100\,({\rm \mu m/s})$, which is not shown, is similar to that in Fig.  \ref{fig-9}(a), whereas configurations similar to that in Fig.  \ref{fig-9}(b) emerge when $V_e$ is increased to $V_e\!=\!150\,({\rm \mu m/s})$. This suggests that the conformally flat configuration emerges at relatively high fluid velocities and that the inclusion of $\xi H_K$ in the Hamiltonian used for the MC update of $\vec{\tau}$ reduces the velocity of faster fluid flows, as shown in Figs. \ref{fig-5}(c) and \ref{fig-7}(b).

The problem that should be clarified is why the conformally flat configuration of $\vec{\tau}$ is compatible with the velocity profile of Couette flow, which is governed by the viscous term in the LNS equation. The viscous term corresponds to the viscous Hamiltonian $H_V$ in Eq. (\ref{discrete-HV}). To examine the compatibility between  $\vec{\tau}$ and $\vec{V}$, the relative mass $m(\theta)$ in Eq. (\ref{relative-mass-theta}) and the relative viscosity ${\boldsymbol \varGamma}(\theta)$ in Eq. (\ref{relative-viscosity-TH}) are plotted in Figs. \ref{fig-9}(c) and \ref{fig-9}(d), respectively. The position of $\theta$ at which $m(\theta)$ reaches its minimum is shown  by the symbol ($\textcolor{red}{\bigcirc}$) in Fig. \ref{fig-9}(c), and ${\boldsymbol \varGamma}(\theta)$ reaches its minimum at nearly the same values of $\theta$ in Fig. \ref{fig-9}(d). Thus, this consistency suggests that the flow field is controlled by the viscous term associated with the viscous Hamiltonian $H_V$ and energetically favors the same $\vec{\tau}$ structure as that expected from $H_K$.

%%%%%%%%%%%%%%%% f-10
\begin{figure}[h!]
	\centering{}\includegraphics[width=10.0cm,clip]{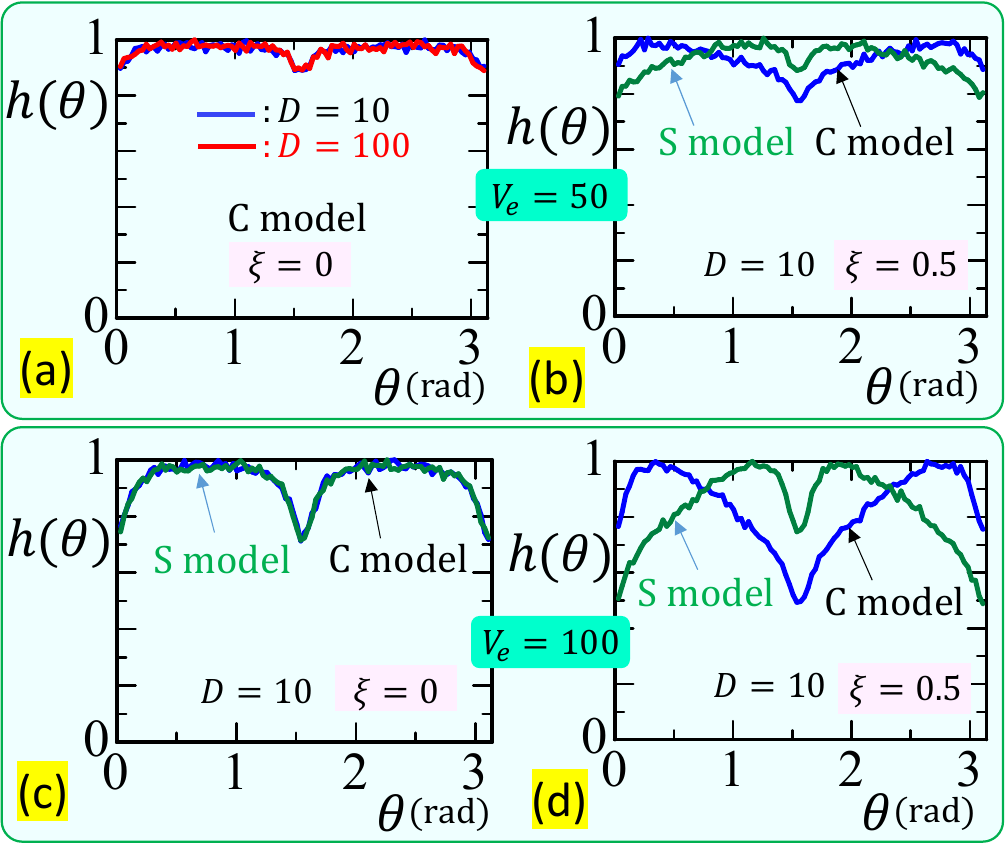}
	\caption{
		Numerically obtained normalized distribution $h(\theta)$  vs. $\theta$ for (a) $\xi\!=\!0$ with $D\!=\!10$ and $D\!=\!100$, and for (b) $\xi\!=\!0.5$, under $V_e\!=\!50\,({\rm \mu m/s})$. The data obtained under  $V_e\!=\!100\,({\rm \mu m/s})$ are plotted in (c) and (d). For $\xi\!=\!0$, the C and S models give identical results as shown in (c), whereas for $\xi\!=\!0.5$, they give slightly different results as shown in (b),(d).
		\label{fig-10}}
\end{figure}
%%%%%%%%%%%%%%%
To confirm this consistency, we calculate the numerically obtained distributions of $\theta$, denoted by $h(\theta)$; 
\begin{eqnarray}
	\label{distribution-of-m}
	h(\theta)= {\rm normalized \; distribution\; of}\;\theta.
\end{eqnarray}
 The density $h(\theta)$ is a histogram of $\theta$ over all lattice sites $(i,j)$. If $h(\theta)$ is high at $\theta\!=\!\theta_0$, this indicates that $\vec{\tau}$ takes the value $\theta\!=\!\theta_0$ at many sites $(i,j)$. For $V_e\!=\!50 ({\rm \mu m /s})$, we find two plateau-like peaks in $h(\theta)$ in Figs. \ref{fig-10}(a), (b), at $\theta\!\simeq\! \pi/4$ and $\theta\!\simeq\! 3\pi/4$. The values of $h(\theta)$ at these peaks are larger than those at $\theta\!\simeq\! 0,\pi$ and $\theta\!\simeq\! \pi/2$; however, the difference is not always apparent.  
  This behavior of $h(\theta)$ with respect to the $\theta$ variation is more prominent when $V_e$ is increased to $V_e\!=\!100$, as shown in Figs. \ref{fig-10}(c) and \ref{fig-10}(d). 
This plateau behavior in $h(\theta)$ for $\xi\!=\!0$  (Fig. \ref{fig-10}(c)) implies that the configuration, close to $\theta\!=\!\pi/4$ and $\theta\!=\!3\pi/4$, represents the minimum-energy state of $H_V$ under the isotropic condition in Eq. (\ref{no-anisotropy}), as confirmed by  the minimum of ${\boldsymbol \varGamma}(\theta)$ in Fig. \ref{fig-9}(d). In contrast, the peak position in $h(\theta)$ for $\xi\!=\!0.5$ (Fig. \ref{fig-10}(d), C model) slightly moves toward $\theta\!=\!0$ and $\theta\!=\!\pi$. This is due to the asymmetry in the effects of ${\boldsymbol \varGamma}_x$ and ${\boldsymbol \varGamma}_y$ in Eq. (\ref{relative-viscosity-TH}) on the energy $H_V$. In the boundary region of the C model, the influence of ${\boldsymbol \varGamma}_y$ is more prominent than that of ${\boldsymbol \varGamma}_x$ because ${\boldsymbol \varGamma}_y$ couples to a nonzero velocity difference, such as $(\vec{V}_2\!-\!\vec{V}_i)^2$, whereas ${\boldsymbol \varGamma}_x$ couples to $(\vec{V}_1\!-\!\vec{V}_i)^2$, which is considerably smaller than $(\vec{V}_2\!-\!\vec{V}_i)^2$ in the boundary region. The peak position of the S model shifts closer to $\theta\!=\!\pi/2$ because  ${\boldsymbol \varGamma}_x$ and ${\boldsymbol \varGamma}_y$ are interchanged in the S model (Appendix \ref{App-A}). This asymmetry between ${\boldsymbol \varGamma}_x$ and ${\boldsymbol \varGamma}_y$ is not reflected in $h(\theta)$ for either the C or S model in Fig. \ref{fig-10}(c) for  $\xi\!=\!0$, because $\xi H_V$ is not included in the total Hamiltonian when $\xi\!=\!0$.

%%%%%%%%%%%%%%%% f-11
\begin{figure}[h!]
	\centering{}\includegraphics[width=10.0cm,clip]{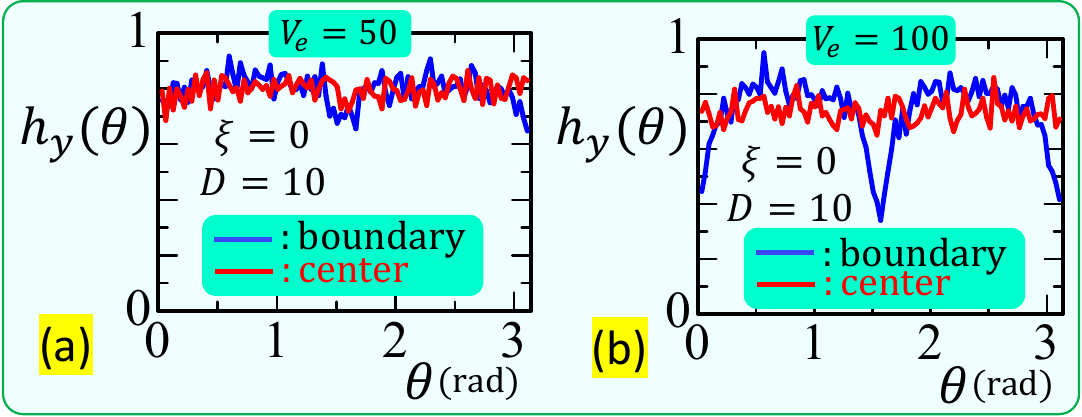}
	\caption{
		Numerically obtained normalized distribution $h_y(\theta)$ vs. $\theta$ for  $\xi\!=\!0$ with $D\!=\!10$ under (a) $V_e\!=\!50\,({\rm \mu m/s})$ and (b) $V_e\!=\!100\,({\rm \mu m/s})$.
		\label{fig-11}}
\end{figure}
%%%%%%%%%%%%%%%
 To further clarify the correspondence between  the minimum energy states of $H_K$ and $H_V$, we define $y$-dependent distribution of $\theta$ such that
  \begin{eqnarray}
 	\label{distribution-of-m-y}
 	h_y(\theta)= {\rm normalized \; distribution\; of}\; \theta\; {\rm at}\; y, 
 \end{eqnarray}
which is a histogram of $\theta$ at lattice sites $(i,j)$ with fixed $j(=\!y)$. For the normalization, the peak value is identified from the data $\{h_y(\theta),1\!\leq\!y\!\leq\!40,0\!\leq\!\theta\!\leq\!\pi)\}$.
We  plot the $h_y(\theta)$ distributions in Figs. \ref{fig-11}(a),(b) with blue and red lines, denoted by 'boundary' and 'center', respectively, defined by
\begin{eqnarray}
	\label{distribution-of-m-y-BR}
	\begin{split}
		&h_y(\theta)({\rm boundary})= \frac{1}{2}\left[h_y(\theta)(j\!=\!3)+h_y(\theta)(j\!=\!38)\right], \; (38=n_y-2),\\
		&h_y(\theta)({\rm center})= \frac{1}{2}\left[h_y(\theta)(j\!=\!20)+h_y(\theta)(j\!=\!21)\right],
	\end{split}
\end{eqnarray}
where the positions denoted by $j\!=\!3$, $j\!=\!38$, and $j\!=\!20$, $j\!=\!21$ are indicated by arrows in Fig. \ref{fig-9}(a). 

We find a clear difference in $h_y(\theta)$ between the boundary and central regions for $V_e\!=\!100\,({\rm \mu m/s})$ in Fig. \ref{fig-11}(b). Sharp valleys in $h_y(\theta)$(boundary) imply that the total number of $\vec{\tau}$ with $\theta\!\simeq\! \pi/4$ and $\theta\!\simeq\! 3\pi/4$  is significantly larger than that with $\theta\!\simeq\! 0,\pi$ and $\theta\!\simeq\! \pi/2$ at the boundary, as confirmed in the snapshot in Fig. \ref{fig-9}(b). The curve at the center is almost constant reflecting that the configuration is almost random isotropic. 

Thus, it is confirmed that the numerically obtained distributions $h_y(\theta)$ are consistent with the expectation from $m(\theta)$, which predicts a conformally flat configuration of $\vec{\tau}$ in the boundary region as an energetically favored configuration for $H_K$. In addition, the conformally flat configuration is also energetically favored with respect to $H_V$ when $\xi\!=\!0$, in which case no difference between the results of the S model and the C model is expected. Even when $\xi H_V$ is included with $\xi\!=\!0.5$, the conformally flat configuration that minimizes $H_K$ remains energetically favored because the energy scale of $H_K$  is comparable to that of $\xi H_V$, which is also minimized by the conformally flat configuration under the isotropic condition in Eq. (\ref{no-anisotropy}). 
Therefore, we find that the spatial variation of the flow field $\vec {V}$ is reflected in the viscosity through a conformal geometric property of the $\vec{\tau}$ configuration.

We should note that the $\vec{\tau}$ configuration is reflected neither in the standard order parameter defined by
\begin{eqnarray}
	\label{order-prm}
	q_\mu=2 \left(\frac{1}{N}\sum_{i=1}^{N}\left[\tau^\mu_{i}\right]^2-0.5\right), \quad (\mu=x,y)
\end{eqnarray}
nor in the $y$-dependent order parameter defined by
\begin{eqnarray}
	\label{order-prm-y}
	q_\mu(y)=2 \left(\frac{1}{n_x}\sum_{i=1}^{n_x}\left[\tau^\mu_{i}(y)\right]^2-0.5\right), \quad (\mu=x,y).
\end{eqnarray}
In fact, we find that $q_\mu\!\simeq\!0$ and $q_\mu(y)\!\simeq\!0$, which are not plotted, implying that the configurations are randomly aligned or that there is no specific alignment direction, as confirmed by the snapshot in Figs. \ref{fig-9}(a),(b).

%%%%%%%%%%%%%%%
\subsection{Mechanism of viscosity reduction %in the boundary region
	\label{viscosity-reduction}}
%%%%%%%%%%%%%%%%
%%%%%%%%%%%%%%%% f-12
\begin{figure}[h!]
	\centering{}\includegraphics[width=10.0cm,clip]{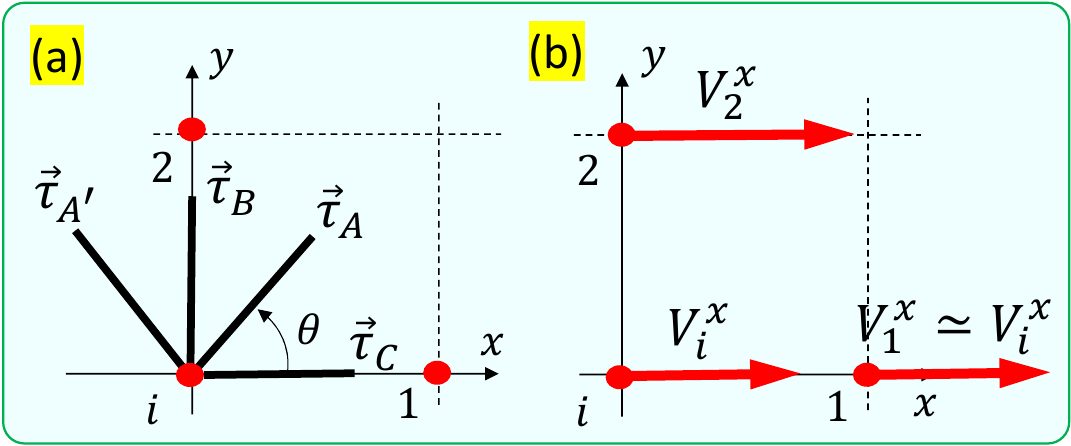}
	\caption{
		(a) Conformally flat ($\vec{\tau}_A$,$\vec{\tau}_{A^\prime}$) and non-flat ($\vec{\tau}_B$, $\vec{\tau}_C$) configurations, and (b) flow field configurations at the sites $i$, 1 and 2 in the upper boundary region.
		\label{fig-12}}
\end{figure}
%%%%%%%%%%%%%%%
In this subsection, we discuss in detail why the relative viscosity decreases in the boundary region (Fig. \ref{fig-6}(a)).
In Fig. \ref{fig-12}(a), four $\vec{\tau}$s at $i$ are plotted; $\vec{\tau}_{A,A^\prime}$ are conformally flat and $\vec{\tau}_{B,C}$ are conformally non-flat. Here, we assume the same configurations of $\vec{\tau}$ at the sites 1 and 2. 
To discuss the viscosity reduction in the boundary region, we assume  that the configurations $\vec{\tau}_{A,A^\prime}$ are dominant there, like in the snapshot in  Fig. \ref{fig-9}(b). 

First,  for the C model, we evaluate $m_i(\tau)$ appearing in $H_K\!=\!(1/2)\sum_i m_i(\tau)\vec{V}_i\cdot\vec{V}_i$ as follows:
\begin{eqnarray}
	\label{HK-ABC}
	\begin{split}
		m_{i}(\tau_A)&=\frac{{\bar m}^{-1}}{\chi_{i1}(\tau_A)\chi_{i2}(\tau_A)}=\frac{{\bar m}^{-1}}{(\cos^2\frac{\pi}{4}+\chi_0^m)(\sin^2 \frac{\pi}{4}+\chi_0^m)}=\frac{{\bar m}^{-1}}{(0.501)^2}\simeq 0.204,\\
		m_{i}(\tau_B)&=\frac{{\bar m}^{-1}}{\chi_{i1}(\tau_B)\chi_{i2}(\tau_B)}=\frac{{\bar m}^{-1}}{(\cos^2\frac{\pi}{2}+\chi_0^m)(\sin^2 \frac{\pi}{2}+\chi_0^m)}=\frac{{\bar m}^{-1}}{0.01\times 1.01}\simeq 5.07,\\
		m_{i}(\tau_C)&=\frac{{\bar m}^{-1}}{\chi_{i1}(\tau_C)\chi_{i2}(\tau_C)}=\frac{{\bar m}^{-1}}{(\cos^20+\chi_0^m)(\sin^2 0+\chi_0^m)}=\frac{{\bar m}^{-1}}{1.01\times0.01}\simeq 5.07,
	\end{split}
\end{eqnarray}
where $m_i(\vec{\tau}_{A^\prime})(=\!m_i(\vec{\tau}_{A}))$ is not listed, for simplicity. The expression $m(\theta)$ in Eq. (\ref{relative-mass-theta}) is used for these evaluations. Under the assumption that $\vec{\tau}_{A,A^\prime}$ are dominant, the evaluations of $m_{i}(\tau_B)$ and $m_{i}(\tau_C)$ are unnecessary; however, they are included for reference.  
We find from these values that $m_i(\tau_A)$ is smaller than the others, in particular, than that for the randomly isotropic $\vec{\tau}_{\rm iso}$ shown in Fig. \ref{fig-8}(b):
\begin{eqnarray}
	\label{flat-vs-isotropic-for-m}
	m_i(\tau_{A})\simeq 0.204 < m_i (\tau_{\rm iso})\simeq 1.
\end{eqnarray}
This implies that the configuration $\tau_A$ is energetically favored, at least in the boundary region, where the extensive part $\vec{V}_i\cdot\vec{V}_i$ of $H_K$ is nonzero and finite and is larger than that in the central region. We note that this result is independent of whether the model is C or S in Eq. (\ref{discrete-chi-for-HK-main}).

It is also useful to evaluate $\Gamma_{i1}$ and $\Gamma_{i2}$ in $H_V$ in the boundary region. It is sufficient to evaluate $\Gamma_{i2}$ because $\Gamma_{i1}(\vec{V}_1\!-\!\vec{V}_i)^2\!\simeq\!0$, whereas $\Gamma_{i2}(\vec{V}_2\!-\!\vec{V}_i)^2\!\not=\!0$ in $H_V$, due to the velocity configuration shown in Fig. \ref{fig-12}(b). Under the assumed configurations of $\vec{\tau}$, $\frac{\chi_{i2}}{\chi_{i1}}\!=\!\frac{\chi_{2i}}{\chi_{25}}$ holds, and we find that
\begin{eqnarray}
	\label{HV-ABC-C-model}
	\begin{split}
	\Gamma_{i2}(\tau_A)&={\bar \Gamma}^{-1}2\frac{\chi_{i2}(\tau_A)}{\chi_{i1}(\tau_A)}={\bar \Gamma}^{-1}2\frac{\sin^2(\frac{\pi}{4})+\chi_0^v}{\cos^2 (\frac{\pi}{4})+\chi_0^v}={\bar \Gamma}^{-1}2\frac{0.5+0.5}{0.5+0.5}\simeq 0.763\; ({\rm C\; model}),\\
	\Gamma_{i2}(\tau_B)&={\bar \Gamma}^{-1}2\frac{\chi_{i2}(\tau_B)}{\chi_{i1}(\tau_B)}={\bar \Gamma}^{-1}2\frac{\sin^2(\frac{\pi}{2})+\chi_0^v}{\cos^2 (\frac{\pi}{2})+\chi_0^v}={\bar \Gamma}^{-1}2\frac{1+0.5}{0.5}\simeq 2.29\; ({\rm C\; model}),\\
	\Gamma_{i2}(\tau_C)&={\bar \Gamma}^{-1}2\frac{\chi_{i2}(\tau_C)}{\chi_{i1}(\tau_C)}={\bar \Gamma}^{-1}2\frac{\sin^2(0)+\chi_0^v}{\cos^2 (0)+\chi_0^v}={\bar \Gamma}^{-1}2\frac{0.5}{1+0.5}\simeq 0.254\; ({\rm C\; model}),\\
	\end{split}
\end{eqnarray}
where the expression ${\boldsymbol \varGamma}_y(\theta)$ in Eq. (\ref{relative-viscosity-TH}) is used to evaluate these specific $\vec{\tau}$s. For the S model, we find that 
\begin{eqnarray}
	\label{HV-ABC-S-model}
	\begin{split}
		\Gamma_{i2}(\tau_A)&={\bar \Gamma}^{-1}2\frac{\chi_{i2}(\tau_A)}{\chi_{i1}(\tau_A)}={\bar \Gamma}^{-1}2\frac{\cos^2(\frac{\pi}{4})+\chi_0^v}{\sin^2 (\frac{\pi}{4})+\chi_0^v}={\bar \Gamma}^{-1}2\frac{0.5+0.5}{0.5+0.5}\simeq 0.763\; ({\rm S\; model}),\\
		\Gamma_{i2}(\tau_B)&={\bar \Gamma}^{-1}2\frac{\chi_{i2}(\tau_B)}{\chi_{i1}(\tau_B)}={\bar \Gamma}^{-1}2\frac{\cos^2(\frac{\pi}{2})+\chi_0^v}{\sin^2 (\frac{\pi}{2})+\chi_0^v}={\bar \Gamma}^{-1}2\frac{0.5}{1+0.5}\simeq 0.254\; ({\rm S\; model}),\\
		\Gamma_{i2}(\tau_C)&={\bar \Gamma}^{-1}2\frac{\chi_{i2}(\tau_C)}{\chi_{i1}(\tau_C)}={\bar \Gamma}^{-1}2\frac{\cos^2(0)+\chi_0^v}{\sin^2 (0)+\chi_0^v}={\bar \Gamma}^{-1}2\frac{1+0.5}{0.5}\simeq 2.29\; ({\rm S\; model}).\\
	\end{split}
\end{eqnarray}
Comparing the results in Eqs. (\ref{HV-ABC-C-model}) and (\ref{HV-ABC-S-model}), we find that only the configuration $\tau_A$ is common to the C and S models. Thus, this value of $\Gamma_{i2}(\tau_A)\!=\!0.763$ is expected in the boundary region in both the C and S models. In contrast, in the central region, both extensive parts $(\vec{V}_1\!-\!\vec{V}_i)^2$ and $(\vec{V}_2\!-\!\vec{V}_i)^2$ are nearly zero, and therefore, the $\tau$ configuration is not uniquely determined and becomes almost random isotropic. Therefore, it is expected that $\Gamma_{i2}(\tau_{\rm iso})\!\simeq\!1$ in the central region. Thus,  we find that
\begin{eqnarray}
	\label{flat-vs-isotropic-for-Gamma}
\Gamma_{i2}(\tau_{A})\simeq 0.763 < \Gamma_{i2} (\tau_{\rm iso})\simeq 1.
\end{eqnarray}
This inequality explains the reduction in the relative viscosity $\Gamma_{i2}$, which defines $\Gamma_y(y)$ in Eq. (\ref{relative-viscosity}), in the conformally flat configuration.

% Section %%%%%%%%%%%%%%%%%%%%%%%%%%%%%%%%%%%%%%%%%%%%%%%%%%%%%%%%%%%%%%%%%%%%%%%%%
\section{Summary and concluding remarks \label{conclusion}}
%%%%%%%%%%%%%%%%%%%%%%%%%%%%%%%%%%%%%%%%%%%%%%%%%%%%%%%%%%%%%%%%%%%%%%%%%%%%%%%%%%%
We study a model of protoplasmic streaming in plant cells as a two-dimensional (2D) Couette flow field between parallel plates using the Langevin and Navier-Stokes equation, which includes a Gaussian random force. An internal degree of freedom (IDOF) $\vec{\tau}$ ($\in S^1/2:$ half-unit circle) corresponding to biological materials is introduced to define position- and direction-dependent viscosity based on Finsler geometry modeling. In our model system, the flow field  is driven by the moving parallel plates, or boundary flows. The supplied energy is dissipated into the fluid through viscosity, and a part of the supplied energy activates the biological materials, causing them to move with the flow field. For this reason, the kinetic energy is included in Hamiltonian for updating the IDOF variable $\vec{\tau}$ by Monte Carlo procedure.

The Reynolds number is very low: $R_e\!=\!2.5\!\times\! 10^{-3}$ or $R_e\!=\!2.5\!\times\! 10^{-4}$, and the advection term is negligible, although it is included for completeness. The parameter $D$ of the Gaussian random force is determined from the kinematic viscosity of protoplasmic streaming using the fluctuation-dissipation relation of Landau-Lifschitz. The boundary velocity is fixed at $V_e(=\!50 ({\rm \mu m /s}))$ in the simulations, and it is also varied over the range $10\!\leq\! V_e\!\leq\! 150 ({\rm \mu m /s})$ to investigate the dependence of the relative shear stress  and other quantities on the boundary velocity.

We find a linear velocity profile, $V_x\!\propto\! y$. The linear behavior of $V_x$ in the Couette flow is not influenced by the IDOF under the assumed condition for the streaming. In contrast to this linear behavior of $V_x$, the relative viscosity $\Gamma^y(y)$ depends on $y$ and reaches a minimum near the boundary. Consequently, the relative viscosity $\Gamma^y_B(y)$ at the boundary decreases with increasing shear rate $\frac{dV_x}{dy}$. This indicates that shear-thinning behavior, which is typical of non-Newtonian fluids, is observed when $V_e$ is varied in the range $V_e\!\geq\!50 ({\rm \mu m /s})$.

Moreover, the directional alignment of the IDOF and its $y$ dependence are analyzed by examining the $\theta$ dependence of $\vec{\tau}$, where $\theta$ is the angle between $\vec{\tau}$ and the $x$ axis. We find that the distribution of $\vec{\tau}$ has peaks at $\theta\!\simeq\! \pi/4$ and $\theta\!\simeq\! 3\pi/4$, resulting in a Finsler metric that is conformally flat, or equivalently, conformally equivalent to the Euclidean metric. This peak structure arises from a reduction in the relative mass $m(\theta)$ in the kinetic energy Hamiltonian near the boundary. A reduction in the relative viscosity ${\bf \varGamma}(\theta)$ is also expected  and observed for these conformally flat configurations near the boundary. 
These analyses reveal that the tendency of the $\vec{\tau}$ distribution toward conformally flat configurations is connected to the rheological properties and, in particular, characterizes the shear-thinning behavior of two-dimensional Couette flow as a model of protoplasmic streaming.

Since the $\theta$ dependence of $m(\theta)$ is determined by the definition of the Finsler metric as shown in Appendix \ref{App-C}, our modeling technique can be used to study the rheological properties of non-Newtonian fluids in Couette flows without Brownian forces.  Shear thinning and shear thickening of non-Newtonian fluids at higher Reynolds numbers, where the advection term is not negligible, will be studied in future work.

\begin{acknowledgments}
The author HK would like to thank Tetsuya Uchimoto for helpful discussions during the early stages of this study. \end{acknowledgments}

\appendix

% App A
%%%%%%%%%%%%%
\section{Discretization of viscous energy Hamiltonian and Laplace operator \label{App-A}}
\begin{figure}[h]
	\centering{}\includegraphics[width=11.0cm,clip]{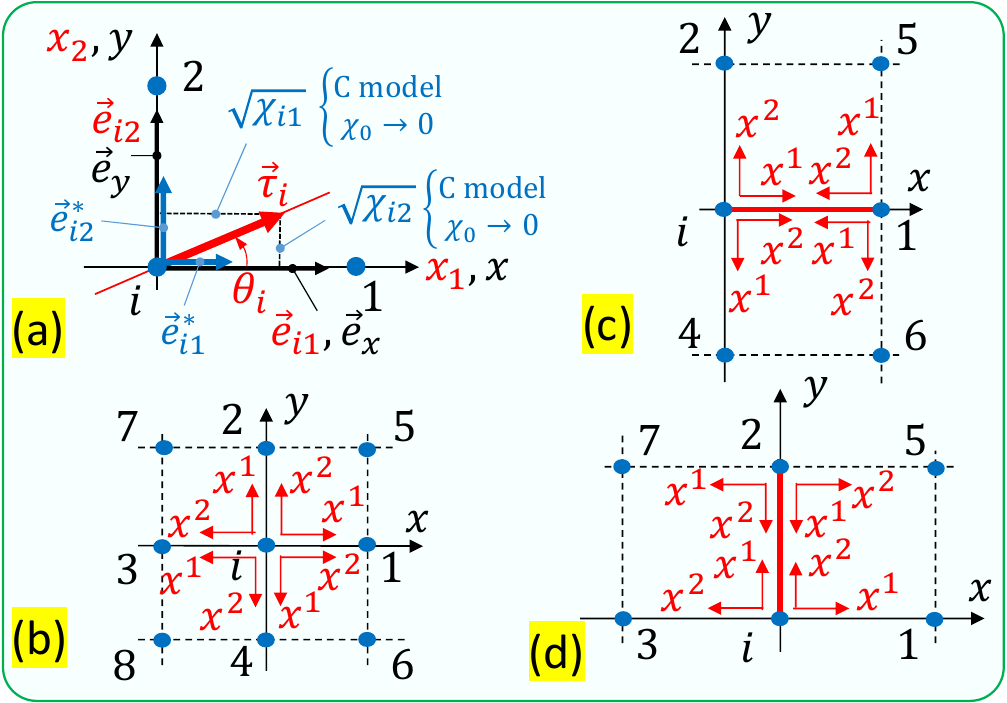}
	\caption{
	(a) The canonical basis vectors $\vec{e}_\mu$, ($\mu\!=\!x,y$) of the Cartesian coordinate axes $(x,y)$, and the unit vectors $\vec{e}_{ij}$, ($j\!=\!1,2$) and covariant basis vectors $\vec{e}_{ij}^{\,*}$, ($j\!=\!1,2$) of the local coordinates $(x^1,x^2)$ at vertex $i$. The vectors $\vec{e}_\mu$, ($\mu\!=\!x,y$) are independent of vertex $i$, whereas $\vec{e}_{ij}$, ($j\!=\!1,2$) and $\vec{e}_{ij}^{\,*}$, ($j\!=\!1,2$) change depending on the choice of local coordinate axes. The correspondence between the lengths of the dashed lines and $\chi_{ij}, (j=1,2)$ is reversed when the C model is replaced by the S model.   Four possible local coordinate systems are shown: (b) at vertex $i$, (c) along bond $i1$, and (d) along bond $i2$. 
		\label{fig-A-1}}
\end{figure}

%%%%%%%%%%%%%
The continuous form of the viscous energy is given by
\begin{eqnarray}
	\label{viscous-cont-Hamiltonian}
%\begin{split}
   H_V=\int \sqrt{g}d^2 xg^{ij}\frac{\partial \vec{V}}{\partial x^i}\cdot \frac{\partial \vec{V}}{\partial x^j}.
%\end{split}
\end{eqnarray}
The Finsler metric is a $2$ by $2$ matrix 
\begin{eqnarray}
	\label{F-metric-2D}
	\begin{split}
		&g_{ij}=\begin{pmatrix}
			(\chi_{i1})^{-2} & 0 \\
			0 & (\chi_{i2})^{-2}  \\
		\end{pmatrix},\\
		&\sqrt{g}=\sqrt{\det g_{ij}}=\frac{1
		}{\chi_{i1}\chi_{i2}}, \quad g^{ij}=(g_{ij})^{-1},
	\end{split}
\end{eqnarray}
where $\chi_{i1}$ and $\chi_{i2}$  are defined on the local coordinate axes $x^1$ and $x^2$ on a square lattice (Figs. \ref{fig-A-1}(a)--(d)) such that
\begin{eqnarray}
	\label{discrete-chi-for-HV}
	\begin{split}
		\chi_{i1}&=\left\{ \begin{array}{@{\,}ll}
			\quad|\vec{\tau}_i\cdot\vec{e}_{i1}|^2+\chi_0^v\;\;\;=\cos^2\theta_{i1}+\chi_0^v\quad  ({\rm C\; model})  \\
			1-(\vec{\tau}_i\cdot\vec{e}_{i1})^2+\chi_0^v=\sin^2\theta_{i1}+\chi_0^v\quad  ({\rm S\;model})
		\end{array} \right.,\\
		\chi_{i2}&=\left\{ \begin{array}{@{\,}ll}
	 \quad|\vec{\tau}_i\cdot\vec{e}_{i2}|^2+\chi_0^v\;\;\;=\cos^2\theta_{i2}+\chi_0^v\quad  ({\rm C\; model}) \\
	1-(\vec{\tau}_i\cdot\vec{e}_{i2})^2+\chi_0^v=\sin^2\theta_{i2}+\chi_0^v\quad  ({\rm S\;model})
\end{array} \right., \\ \chi_0^v&=0.5.
	\end{split}
\end{eqnarray}
The symbols $\theta_{i1}$ and $\theta_{i2}$ denote the angles between $\vec{\tau}_i$ and $\vec{e}_{i1}$ and between $\vec{\tau}_i$ and $\vec{e}_{i2}$, respectively. We should note that $\theta_{i1}\!=\!\theta_i$, where $\theta_i$ is the angle between $\vec{\tau}$ and the $x$ axis as shown in Fig. \ref{fig-A-1}(a).

%%%%%%%%%%%%%%%% f-A-1
%%%%%%%%%%%%%%%

To define the Hamiltonian for viscous energy, we use both Cartesian and local coordinate systems on ${\bf R}^2$. Let $\vec{e}_\mu, (\mu\!=\!x,y)$ denote the basis vectors of the Cartesian coordinate system, and let $\vec{e}_{ij}, (j\!=\!1,2)$ denote the unit vectors pointing from vertex $i$ to vertex $j$ (Fig. \ref{fig-A-1}(a)). We note that the covariant vectors $\vec{e}_{ij}^{\,*}, (j\!=\!1,2)$, which form the basis vectors of the local coordinate system (Fig. \ref{fig-A-1}(a)), are defined as follows:
\begin{eqnarray}
	\label{covariant-basis-vectors}
	\begin{split}
		\vec{e}_{i1}^{\,*} = \frac{1}{\chi_{i1}}\vec{e}_{i1},\quad \vec{e}_{i2}^{\,*} = \frac{1}{\chi_{i2}}\vec{e}_{i2}.
	\end{split}
\end{eqnarray}
It is straightforward to verify that $g_{11}\!=\!\vec{e}_{i1}^{\,*} \cdot \vec{e}_{i1}^{\,*}$, $g_{22}\!=\!\vec{e}_{i2}^{\,*} \cdot \vec{e}_{i2}^{\,*}$, and $g_{12}\!=\!g_{21}\!=\!0$. 

We can also define a metric tensor $g_{ij}^\#$ such that
\begin{eqnarray}
	\label{new-F-metric-2D}
	\begin{split}
		g_{ij}^\#=\begin{pmatrix}
			\chi_{i2}^{2} & 0 \\
			0 & \chi_{i1}^{2}  \\
		\end{pmatrix},
	\end{split}
\end{eqnarray}
which is conformally equivalent to $g_{ij}$ given in Eq. (\ref{F-metric-2D}). The equivalence is checked as follows:
\begin{eqnarray}
	\label{F-metric-2D-conf}
	\begin{split}
		g_{ij}^\#=\begin{pmatrix}
			\chi_{i2}^{2} & 0 \\
			0 & \chi_{i1}^{2}  \\
		\end{pmatrix}= \chi_{i1}^{2}\chi_{i2}^{2}\begin{pmatrix}
			\chi_{i1}^{-2} & 0 \\
			0 & \chi_{i2}^{-2}  \\
		\end{pmatrix}\simeq \begin{pmatrix}
			\chi_{i1}^{-2} & 0 \\
			0 & \chi_{i2}^{-2}  \\
		\end{pmatrix}=g_{ij}.
	\end{split}
\end{eqnarray}
Note that the conformal transformation $\chi_{i1}^{2}\chi_{i2}^{2}g_{ij}$ of $g_{ij}$ applies to the interaction scale, not to the spatial length scale \cite{Andersen-etal-NatPhys2025}. The length scale in space is fixed at the Euclidean scale of ${\bf R}^2$, and no symmetry transformation is applied to the spatial coordinates. 
In this new gauge defined in Eq. (\ref{new-F-metric-2D}), the basis vectors are given by $\vec{e}_{i1}^{\,\#} \!=\! {\chi_{i2}}\vec{e}_{i1},\; \vec{e}_{i2}^{\,\#} \!=\! {\chi_{i1}}\vec{e}_{i2}$. It is straightforward to verify that $H_V(g_{ij})\!=\!H_V(g_{ij}^\#)$ in Eq. (\ref{viscous-cont-Hamiltonian}). This conformal invariance of $H_V$ is characteristic of two-dimensional models, regardless of whether the symmetry applies to the interaction strength or to spatial transformations \cite{Andersen-etal-NatPhys2025}.

The flow field $\vec{V}_i$ at vertex $i$ can be expressed as $\vec{V}_i\!=\!V_i^x\vec{e}_x\!+\!V_i^y\vec{e}_y\!=\!(V_i^x,V_i^y)$. The differentials  $\nabla \vec{V}_i\cdot\nabla \vec{V}_i$  are expressed as differences using the local coordinates such that 
$g^{11}\frac{\partial \vec{V}_i}{\partial x^1} \cdot \frac{\partial \vec{V}_i}{\partial x^1} +g^{22}\frac{\partial \vec{V}_i}{\partial x^2} \cdot \frac{\partial \vec{V}_i}{\partial x^2}\to g^{11}(\vec{V}_1\!-\!\vec{V}_i)^2\!+\! g^{22}(\vec{V}_2\!-\!\vec{V}_i)\!=\!\chi_{i1}^{2}(\vec{V}_1\!-\!\vec{V}_i)^2\!+\!\chi_{i2}^{2}(\vec{V}_2\!-\!\vec{V}_i)^2$. Replacing $\int \sqrt{g}d^2x$ with $\sum_i (\chi_{i1}\chi_{i2})^{-1}$ in the expressions of the original gauge in Eq. (\ref{F-metric-2D}), we find 
\begin{eqnarray}
	\label{discrete-HV-1}
H_V(i12)= \frac{\chi_{i1}}{\chi_{i2}}(\vec{V}_1-\vec{V}_i)^2+\frac{\chi_{i2}}{\chi_{i1}}(\vec{V}_2-\vec{V}_i)^2, 
\end{eqnarray}
for the local coordinate system $(i12)$, where $i$ denotes the vertex index of the origin of the local coordinate system, and $1$ and $2$ denote the nearest-neighbor vertex numbers along the $x_1$ and $x_2$ axes, respectively. Expressions for the other three possible local coordinate systems, $(i23)$, $(i34)$, and $(i41)$ (Fig. \ref{fig-A-1}(b)), are given by
\begin{eqnarray}
	\label{discrete-HV-234}
	\begin{split}
	H_V(i23)&=\frac{\chi_{i2}}{\chi_{i3}}(\vec{V}_2-\vec{V}_i)^2+\frac{\chi_{i3}}{\chi_{i2}}(\vec{V}_3-\vec{V}_i)^2, \\
	H_V(i34)&=\frac{\chi_{i3}}{\chi_{i4}}(\vec{V}_3-\vec{V}_i)^2+\frac{\chi_{i4}}{\chi_{i3}}(\vec{V}_4-\vec{V}_i)^2, \\
	H_V(i41)&=\frac{\chi_{i4}}{\chi_{i1}}(\vec{V}_4-\vec{V}_i)^2+\frac{\chi_{i1}}{\chi_{i4}}(\vec{V}_1-\vec{V}_i)^2.
	\end{split}
\end{eqnarray}
Using the relations $\chi_{i1}\!=\!\chi_{i3}$, $\chi_{i2}\!=\!\chi_{i4}$ and summing over these four terms, we have 
\begin{eqnarray}
	\label{discrete-HV-sum-1}
	\begin{split}
H_V=2\sum_i \left(\frac{\chi_{i1}}{\chi_{i2}}\left[(\vec{V}_1-\vec{V}_i)^2+(\vec{V}_3-\vec{V}_i)^2\right]+\frac{\chi_{i2}}{\chi_{i1}}\left[(\vec{V}_2-\vec{V}_i)^2+(\vec{V}_4-\vec{V}_i)^2\right]\right).
	\end{split}
\end{eqnarray}
The sum over vertices $i$ can be changed to the sum over bonds $ij$ as follows: 
In these terms in the sum $\sum_i$, we find the terms  proportional to $(\vec{V}_1\!-\!\vec{V}_i)^2$ at local coordinates $(i12)$ and $(15i)$  (Fig. \ref{fig-A-1}(c))   are given by $\sqrt{g}g^{11}\frac{\partial \vec{V}_i}{\partial x^1} \cdot \frac{\partial \vec{V}_i}{\partial x^1}\!=\!\frac{\chi_{i1}}{\chi_{i2}}(\vec{V}_1\!-\!\vec{V}_i)^2$ and $\sqrt{g}g^{22}\frac{\partial \vec{V}_1}{\partial x^2} \cdot \frac{\partial \vec{V}_1}{\partial x^2}\!=\!\frac{\chi_{1i}}{\chi_{15}}(\vec{V}_1\!-\!\vec{V}_i)^2$, which can be written as $\gamma_{i1}(\vec{V}_1\!-\!\vec{V}_i)^2$, $\gamma_{i1}\!=\!\frac{\chi_{i1}}{\chi_{i2}}\!+\!\frac{\chi_{1i}}{\chi_{15}}$. The contributions from the coordinates $(i41)$ and $(1i6)$ are identical to $\gamma_{i1}$ because $\chi_{i4}\!=\!\chi_{i2}$ and  $\chi_{16}\!=\!\chi_{15}$. For this reason, the numerical factor 2 appears, as in Eq. (\ref{discrete-HV-sum-1}). However, this factor of 2 can be removed by introducing the normalization factor described below. 
The terms proportional to $(\vec{V}_2\!-\!\vec{V}_i)^2$ are  $\sqrt{g}g^{22}\frac{\partial \vec{V}_i}{\partial x^2} \cdot \frac{\partial \vec{V}_i}{\partial x^2}\!=\!\frac{\chi_{i2}}{\chi_{i1}}(\vec{V}_2\!-\!\vec{V}_i)^2$ and $\sqrt{g}g^{11}\frac{\partial \vec{V}_2}{\partial x^1} \cdot \frac{\partial \vec{V}_2}{\partial x^1}\!=\!\frac{\chi_{2i}}{\chi_{25}}(\vec{V}_2\!-\!\vec{V}_i)^2$ at local coordinates $(i12)$  and $(2i5)$  (Fig. \ref{fig-A-1}(d)), which can be written as  $\gamma_{i2}(\vec{V}_2\!-\!\vec{V}_i)^2$, $\gamma_{i2}\!=\!\frac{\chi_{i2}}{\chi_{i1}}\!+\!\frac{\chi_{2i}}{\chi_{25}}$. Thus, we obtain 
\begin{eqnarray}
	\label{discrete-HV-org}
	H_V= \sum_{i}\left[\Gamma_{i1}\left(\vec{V}_{1}-\vec{V}_{i}\right)^2+\Gamma_{i2}\left(\vec{V}_{2}-\vec{V}_{i}\right)^2\right], 
\end{eqnarray}
which can be written by using the sum over bonds $ij$ such that
\begin{eqnarray}
	\label{discrete-HV}
	\begin{split}
	H_V&= \sum_{ij}\Gamma_{ij}(\vec{V}_i-\vec{V}_j)^2, \\
	\Gamma_{i1}&=\bar{\Gamma}^{-1}\gamma_{i1}, \quad  \gamma_{i1}=\frac{\chi_{i1}}{\chi_{i2}}+\frac{\chi_{1i}}{\chi_{15}}, \\ \Gamma_{i2}&=\bar{\Gamma}^{-1}\gamma_{i2},\quad \gamma_{i2}=\frac{\chi_{i2}}{\chi_{i1}}+\frac{\chi_{2i}}{\chi_{25}},
	\end{split}
\end{eqnarray}
where $\bar{\Gamma}$ is a normalization factor defined by 
\begin{eqnarray}
	\label{normalization-factor}
	\begin{split}
	\bar{\Gamma}=\langle\gamma_{ij}(\tau^{\rm iso})\rangle=\frac{\sum_{ij}\gamma_{ij}(\tau^{\rm iso}) }{\sum_{ij} 1}\left(\simeq 2.62\; {\rm for}\; \chi_0^v=0.5\right).
	\end{split}
\end{eqnarray}
This factor $\bar{\Gamma}$ guarantees that $\Gamma_{ij}\to 1$ or, equivalently, that $H_V\to\sum_{ij}(\vec{V}_i-\vec{V}_j)^2$ for isotropic $\tau$. The factor $\bar{\Gamma}$ is numerically obtained using $5\!\times\!10^5$ isotropic configurations $\{\vec{\tau}_i^{\rm iso},(i=1,\cdots,N)\}$. This value of $\bar{\Gamma}$ for $\chi_0^v\!=\!0.5$ is identical to the analytical result as shown below. 
 
It should be noted that
\begin{eqnarray}
	\label{reciprocal}
	\Gamma_{ij}=\Gamma_{ji}.
\end{eqnarray}
For example, in the case $ij\!=\!i1$, we find from Eq. (\ref{discrete-HV}) that $\gamma_{1i}\!=\!\frac{\chi_{1i}}{\chi_{15}}\!+\!\frac{\chi_{i1}}{\chi_{i2}}$ by interchanging $i$ and $1$, as well as $2$ and $5$. Vertices 2 and 5 are the vertex numbers in the positive $y$ direction from vertices $i$ and 1, respectively, and therefore, they should also be interchanged when $i$ and $1$ are interchanged. The bond $ij$ is along the $x$ ($y$) axis for $j\!=\!1$ ($j\!=\!2$), and therefore, $\Gamma_{i1}$ ($\Gamma_{i2}$) is considered  the $x$- ($y$-) component of $\Gamma_{ij}$.

 Note also that  $\Gamma_{ij}$ is a two-variables function of the angle $\theta_i$, between $\vec{\tau}_{i}$ and the $x$ axis, and the angle $\theta_j$, between $\vec{\tau}_{j}$ and the $x$ axis. 
 For the C model, we have
 \begin{eqnarray}
 	\label{chi-theta-expressions-Gamma-i1}
 	\begin{split}
 \chi_{i1}&=|\vec{\tau}_i\cdot\vec{e}_{i1}|^2+\chi_0^v=\cos^2\theta_i+\chi_0^v,  \quad \chi_{i2}=|\vec{\tau}_i\cdot\vec{e}_{i2}|^2+\chi_0^v=\sin^2\theta_i+\chi_0^v,\\  \chi_{15}&=|\vec{\tau}_1\cdot\vec{e}_{15}|^2+\chi_0^v=\sin^2\theta_1+\chi_0^v,  \quad \chi_{1i}=|\vec{\tau}_1\cdot\vec{e}_{1i}|^2+\chi_0^v=\cos^2\theta_1+\chi_0^v  	\end{split}
\end{eqnarray}
for $\Gamma_{i1}$, and
\begin{eqnarray}
	\label{chi-theta-expressions-Gamma-i2}
	\begin{split}
		\chi_{i1}&=|\vec{\tau}_i\cdot\vec{e}_{i1}|^2+\chi_0^v=\cos^2\theta_i+\chi_0^v,  \quad \chi_{i2}=|\vec{\tau}_i\cdot\vec{e}_{i2}|^2+\chi_0^v=\sin^2\theta_i+\chi_0^v,\\  \chi_{2i}&=|\vec{\tau}_2\cdot\vec{e}_{2i}|^2+\chi_0^v=\sin^2\theta_2+\chi_0^v,\quad \chi_{25}=|\vec{\tau}_2\cdot\vec{e}_{25}|^2+\chi_0^v=\cos^2\theta_2+\chi_0^v
	\end{split}
\end{eqnarray}
for $\Gamma_{i2}$. 
The expressions for the S model are obtained by interchanging $\sin$ and $\cos$. Here, it should be emphasized that the angle $\theta_i$ is not always identical to the angle between $\vec{\tau}_i$ and $\vec{e}_{ij}$. 

Thus, we obtain 
\begin{eqnarray}
	\label{relative-viscosity-i1-angle}
	\begin{split}
		\Gamma_{i1}(\theta_i,\theta_1)&=\bar{\Gamma}^{-1}\left(\frac{\chi_{i1}}{\chi_{i2}}+\frac{\chi_{1i}}{\chi_{15}}\right)=\left\{ \begin{array}{@{\,}ll}
			 \bar{\Gamma}^{-1}\left(\frac{\cos^2\theta_i+\chi_0^v}{\sin^2\theta_i+\chi_0^v}+\frac{\cos^2\theta_1+\chi_0^v}{\sin^2\theta_1+\chi_0^v}\right),\; ({\rm C\; model})\\
 	\bar{\Gamma}^{-1}\left(\frac{\sin^2\theta_i+\chi_0^v}{\cos^2\theta_i+\chi_0^v}+\frac{\sin^2\theta_1+\chi_0^v}{\cos^2\theta_1+\chi_0^v}\right),\; ({\rm S\; model})
		\end{array} \right.,\\
		\Gamma_{i2}(\theta_i,\theta_2)&=\bar{\Gamma}^{-1}\left(\frac{\chi_{i2}}{\chi_{i1}}+\frac{\chi_{2i}}{\chi_{25}}\right)=\left\{ \begin{array}{@{\,}ll}
	\bar{\Gamma}^{-1}\left(\frac{\sin^2\theta_i+\chi_0^v}{\cos^2\theta_i+\chi_0^v}+\frac{\sin^2\theta_2+\chi_0^v}{\cos^2\theta_2+\chi_0^v}\right),\; ({\rm C\; model}) \\
	\bar{\Gamma}^{-1}\left(\frac{\cos^2\theta_i+\chi_0^v}{\sin^2\theta_i+\chi_0^v}+\frac{\cos^2\theta_2+\chi_0^v}{\sin^2\theta_2+\chi_0^v}\right),\; ({\rm S\; model})
\end{array} \right.,\\
\bar{\Gamma}&=\frac{2}{\pi}\int_0^\pi \frac{\cos^2\theta+\chi_0^v}{\sin^2\theta+\chi_0^v}d\theta=\frac{2}{\pi}\int_0^\pi \frac{\sin^2\theta+\chi_0^v}{\cos^2\theta+\chi_0^v}d\theta\; (\simeq 2.62\; {\rm for}\; \chi_0^v=0.5).
	\end{split}
\end{eqnarray}
The normalization factor $\bar{\Gamma}$  is independent of the model.

From these expressions,  we find that  $\Gamma_{i1}\left(\propto\!\frac{\chi_{i1}}{\chi_{i2}}\!+\!\frac{\chi_{1i}}{\chi_{15}}\!=\!\frac{\cos^2\theta_i\!+\!\chi_0^v}{\sin^2\theta_i\!+\!\chi_0^v}\!+\!\frac{\cos^2\theta_1\!+\!\chi_0^v}{\sin^2\theta_1\!+\!\chi_0^v}\; ({\rm C\; model})\right)$ is symmetric under the interchange of $\theta_i$ and $\theta_1$. This implies that the  dependence of $\Gamma_{i1}$ on site $i$ is identical to its dependence on site $1$ when $\theta_i\!=\!\theta_1$. 
The same argument applies to $\Gamma_{i2}$. Hence, the dependence on site $i$ is sufficient to describe both $\Gamma_{i1}$ and $\Gamma_{i2}$ in terms of the dependence on the angle between $\vec{\tau}$ and the $x$ axis, at least for uniformly aligned configurations.  Thus, the expressions for the site $i$ dependence of $\Gamma_{i1}$ and $\Gamma_{i2}$, corresponding to $\frac{\chi_{i1}}{\chi_{i2}}$ and $\frac{\chi_{i2}}{\chi_{i1}}$, respectively, in Eq. (\ref{discrete-HV}) are calculated for the C model as
\begin{eqnarray}
	\label{relative-viscosity-TH}
	\begin{split}
		{\boldsymbol \varGamma}_x(\theta)&=\bar{\boldsymbol \varGamma }_x^{-1}\left(\frac{\cos^2\theta+\chi_0^v}{\sin^2\theta+\chi_0^v}\right),\;
		\bar{\boldsymbol \varGamma}_x=\frac{1}{\pi}\int_0^\pi \left(\frac{\cos^2\theta+\chi_0^v}{\sin^2\theta+\chi_0^v}\right)d\theta\; 	(\simeq 1.31\; {\rm for}\; \chi_0^v=0.5),\\
		{\boldsymbol \varGamma}_y(\theta)&=\bar{\boldsymbol \varGamma }_y^{-1}\left(\frac{\sin^2\theta+\chi_0^v}{\cos^2\theta+\chi_0^v}\right),\;
\bar{\boldsymbol \varGamma}_y=\frac{1}{\pi}\int_0^\pi \left(\frac{\sin^2\theta+\chi_0^v}{\cos^2\theta+\chi_0^v}\right)d\theta\; (\simeq 1.31\; {\rm for}\; \chi_0^v=0.5),\\
		{\boldsymbol \varGamma}(\theta)&={\boldsymbol \varGamma}_x(\theta)+{\boldsymbol \varGamma}_y(\theta)=\bar{\boldsymbol \varGamma }^{-1}\left(\frac{\cos^2\theta+\chi_0^v}{\sin^2\theta+\chi_0^v}+\frac{\sin^2\theta+\chi_0^v}{\cos^2\theta+\chi_0^v}\right),\\
\bar{\boldsymbol \varGamma}&=\bar{\boldsymbol \varGamma}_x+\bar{\boldsymbol \varGamma}_y=\frac{1}{\pi}\int_0^\pi \left(\frac{\cos^2\theta+\chi_0^v}{\sin^2\theta+\chi_0^v}+\frac{\sin^2\theta+\chi_0^v}{\cos^2\theta+\chi_0^v}\right)d\theta\; (\simeq 2.62\; {\rm for}\; \chi_0^v=0.5).
	\end{split}
\end{eqnarray}
%%%%%%%%%%%%%%%% f-A-2
\begin{figure}[h]
	\centering{}\includegraphics[width=11.0cm,clip]{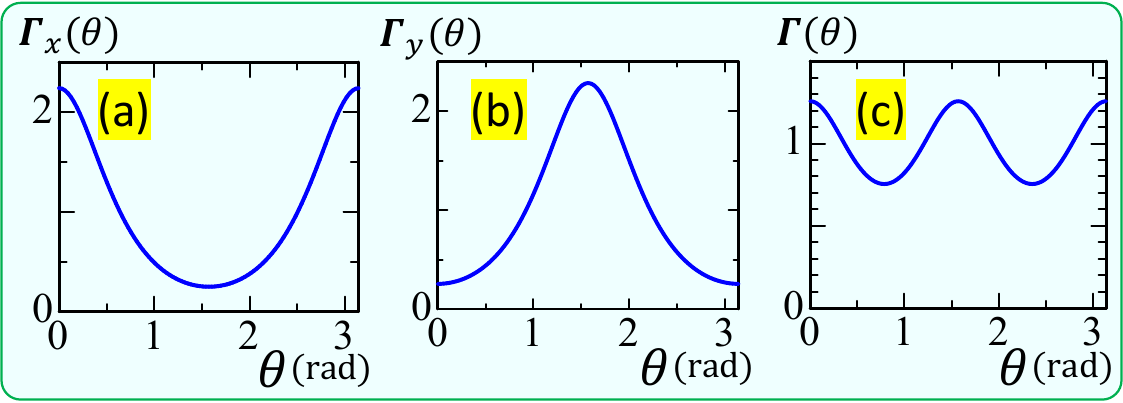}
	\caption{
		(a), (b) The $x$, $y$ components  ${\boldsymbol \varGamma}_x$, ${\boldsymbol \varGamma}_y$ of relative viscosity vs. $\theta$ for C model, and (c) the mean ${\boldsymbol \varGamma}$ of the $x$ and $y$ components. In the S model, ${\boldsymbol \varGamma}_x$ and ${\boldsymbol \varGamma}_y$ are interchanged, while ${\boldsymbol \varGamma}$ remains unchanged. $\chi_0^v$ is fixed at $\chi_0^v\!=\!0.5$.
		\label{fig-A-2}}
\end{figure}
%%%%%%%%%%%%%%%
${\boldsymbol \varGamma}_{x}(\theta)$ and ${\boldsymbol \varGamma}_{y}(\theta)$ represent the site-$i$ dependence of $\Gamma_{i1}$ and  $\Gamma_{i2}$, respectively, and therefore, ${\boldsymbol \varGamma}_{x}(\theta)$ (${\boldsymbol \varGamma}_{y}(\theta)$) is  identical to $\Gamma_{i1}$ ($\Gamma_{i2}$) only when $\vec{\tau}_1\!=\!\vec{\tau}_i$ ($\vec{\tau}_2\!=\!\vec{\tau}_i$). 
${\boldsymbol \varGamma}_x(\theta)$, ${\boldsymbol \varGamma}_y(\theta)$ and ${\boldsymbol \varGamma}(\theta)$ are plotted in Figs. \ref{fig-A-2}(a)--(c).  If $\theta\!\simeq\!\pi/2$, then  ${\boldsymbol \varGamma}_x(\theta)\!\simeq\!0$ and ${\boldsymbol \varGamma}_y(\theta)\!\to\!$ the maximum. On the other hand,  if $\theta\!\simeq\!0$ or $\theta\!\simeq\!\pi$, then  ${\boldsymbol \varGamma}_x(\theta)\!\to\!$  the maximum  and ${\boldsymbol \varGamma}_y(\theta)\!\simeq\!0$. Therefore, anisotropic $\Gamma_{ij}$ is expected when $\vec{\tau}$  uniformly  aligns  along the $x$ or $y$ axis. Such uniformly aligned configurations of  $\vec{\tau}$ are conformally non-flat as shown in Fig. \ref{fig-8}(c). In contrast, for the condition  
\begin{eqnarray}
	\label{no-anisotropy}
{\boldsymbol \varGamma}_x(\theta)\simeq{\boldsymbol \varGamma}_y(\theta),\quad \left(\theta\simeq\frac{\pi}{4}\;{\rm or}\;\theta\simeq\frac{3\pi}{4}\right),
\end{eqnarray}
 the resulting configuration is characterized by the minimum of ${\boldsymbol \varGamma}(\theta)$ and is conformally flat as shown in Fig. \ref{fig-8}(a). 
 
 These properties concerning ${\boldsymbol \varGamma}_{x,y}(\theta)$ and ${\boldsymbol \varGamma}(\theta)$ remain unchanged when $g_{ij}^\#$ in Eq. (\ref{new-F-metric-2D}) is assumed because $H_V$ is conformally invariant. For the same reason, $\Gamma_{ij},(j\!=\!1,2)$ remain unchanged when $g_{ij}^\#$ in Eq. (\ref{new-F-metric-2D}) is used. However, ${m}(\theta)$ does not share this robustness, as shown below in Appendix \ref{App-C}, because $H_K$ is not conformally invariant.

By comparing the continuous and discrete forms of $H_V$ in Eqs. (\ref{viscous-cont-Hamiltonian}) and (\ref{discrete-HV}), respectively, we obtain the discrete Laplace operator
\begin{eqnarray}
	\label{discrete-Laplace}
	\Laplace (\tau) \vec{V}_i=\sum_{j(i)}\Gamma_{ij}(\tau)\left(\vec{V}_j-\vec{V}_i\right)=\sum_{j(i)}\Gamma_{ij}\vec{V}_j-\left(\sum_{j(i)}\Gamma_{ij}\right)\vec{V}_i,
\end{eqnarray}
where $\sum_{j(i)}$ denotes the sum over vertices $j$ connected to vertex $i$. From the second expression, $\Gamma_{ij}$ can naturally be interpreted as a relative viscosity. Additionally, the second expression can also be written as  
\begin{eqnarray}
	\label{relative-Gamma}
%	\begin{split}
\sum_{j(i)}\Gamma_{ij}(\tau)\left(\vec{V}_j-\vec{V}_i\right)=\Gamma_{i1}\left(\vec{V}_1-\vec{V}_i\right)+\Gamma_{i3}\left(\vec{V}_3-\vec{V}_i\right)+\Gamma_{i2}\left(\vec{V}_2-\vec{V}_i\right)+\Gamma_{i4}\left(\vec{V}_4-\vec{V}_i\right).%\\
%&=\Gamma_{i1}\left(\vec{V}_1-\vec{V}_i\right)+\Gamma_{3i}\left(\vec{V}_i-\vec{V}_3\right)+\Gamma_{i2}\left(\vec{V}_2-\vec{V}_i\right)+\Gamma_{4i}\left(\vec{V}_i-\vec{V}_4\right).
%	\end{split}
\end{eqnarray}
Therefore, it is natural to refer to $\Gamma_{ij}$ for a bond $ij$ along the $x$ ($y$) axis, such as $\Gamma_{i1}$ and $\Gamma_{3i}$ ($\Gamma_{i2}$ and $\Gamma_{4i}$), as the $x$ ($y$)-directional component of the relative viscosity.

% App B
%%%%%%%%%%%%%
\section{Fluctuation dissipation relation \label{App-B}}
%%%%%%%%%%%%%
In this Appendix, we discuss the fluctuation-dissipation relation for the 2D Couette flow, following the discussion in Ref. \cite{Noro-etal-PHYSA2023} for 3D cylindrical flow domains, in which IDOF is not assumed.

First, we show the fluctuation-dissipation relation  of the stress tensor $s_{ab}$ 
\begin{eqnarray}
	\label{FDR}
	\left\langle s_{ab} (t,\vec{r})s_{cd}(t^\prime,\vec{r}^{\,\prime})\right\rangle=\frac{2k_BT\rho\nu}{{\rm A}_i}\left(\delta_{ac}\delta_{bd}+\delta_{ad}\delta_{bc}\right)\delta(t-t^\prime),
\end{eqnarray}
where $\delta(\vec{r}\!-\!\vec{r}^{\,\prime})$ is replaced by $1/A_i(=\!1/(\varDelta x)^2)$. 
The stress tensor $s_{ab}$ can be expanded with respect to $\varDelta \vec{x}_b$ such that
\begin{eqnarray}
	\label{stress-tensor}
	\begin{split}
		&s_{ab}(t,\vec{r}+\varDelta \vec{x}_b)=s_{ab}(t,\vec{r})+\frac{\partial s_{ab}}{\partial x_b}(t,\vec{r})\varDelta x,\;(a,b=1,2),\\
		&\varDelta\vec{x}_1=(\varDelta x,0),\; \varDelta\vec{x}_2=(0,\varDelta x).
	\end{split}
\end{eqnarray}
From this, we find
\begin{eqnarray}
	\begin{split}
		\label{FD-A}
		(\varDelta x)^{2}\left\langle \sum_b\frac{\partial s_{ab}}{\partial x_b}(t,\vec{r}) \sum_b\frac{\partial s_{ab}}{\partial x_b}(t^\prime,\vec{r})\right\rangle
		=&\left\langle \sum_bs_{ab} (t,\vec{r}+\varDelta\vec{x}_b)\sum_b s_{ab}(t^\prime,\vec{r}+\varDelta\vec{x}_b)\right\rangle\\
		&+\left\langle \sum_b s_{ab} (t,\vec{r})\sum_b s_{ab}(t^\prime,\vec{r})\right\rangle,
	\end{split}
\end{eqnarray}
where the terms $\left\langle s_{ab} (t,\vec{r})s_{cd}(t^\prime,\vec{r}^{\,\prime}+\varDelta\vec{x}_d^{\,\prime})\right\rangle$ are zero due to $\delta(\vec{r}\!-\!\vec{r}^\prime)$. The first of the right hand side is of the form $\langle (s_{a1}\!+\!s_{a2})(s_{a1}\!+\!s_{a2})\rangle\!=\!\langle s_{a1}s_{a1}\rangle\!+\!\langle s_{a2}s_{a2}\rangle$ for $a\!=\!1,2$. Therefore, from Eq. (\ref{FDR}), we obtain $\left\langle \sum_bs_{ab} (t,\vec{r}\!+\!\varDelta\vec{x}_b)\sum_b s_{ab}(t^\prime,\vec{r}\!+\!\varDelta\vec{x}_b)\right\rangle\!=\!\frac{6k_BT\rho\nu}{{\rm A}_i}\delta(t-t^\prime)$. The second of the right hand side of Eq. (\ref{FD-A}) has the same value, and therefore, the right hand side is $\frac{12k_BT\rho\nu}{{\rm A}_i}\delta(t-t^\prime)$. The left-hand side is identified with $(\varDelta x)^{2}\rho^2\langle \eta_i^a(t) \eta_j^a(t^\prime)\rangle$ from Eq. (\ref{eta-div-of-sab}), and therefore, the left hand side is $(\varDelta x)^{2}\rho^2 2D\delta(t-t^\prime)$ from Eq. (\ref{disc-GRF}). Thus, using ${\rm A}_i\!=\!(\varDelta x)^2$, we obtain
\begin{eqnarray}
	\begin{split}
		\label{FD-B}
		&(\varDelta x)^{2}\rho^2 2D_{\rm fd}\delta(t-t^\prime)=\frac{12k_BT\rho\nu}{{\rm A}_i}\delta(t-t^\prime) \\
		\Leftrightarrow \;&D_{\rm fd}=\frac{6k_BT\nu}{\rho(\varDelta x)^4}.
	\end{split}
\end{eqnarray}
Here, we use the symbol $D_{\rm fd}$ with the suffix 'fd' to denote $D$ in Eq. (\ref{disc-GRF}), since Eq. (\ref{FD-B}) is derived from the fluctuation-dissipation relation by assuming Eq. (\ref{eta-div-of-sab}).
Using the physical parameters in Table \ref{table-1} and $k_BT\!=\!4\times 10^{-21}$, we obtain $D_{\rm fd}\!=\!9.8\times 10^{-9} ({\rm \frac{m^2}{s^3}})\!=\!9.8 ({\rm \frac{{[\alpha m]^2}}{{[\beta s]^{3}}}})$ for case A and $D_{\rm fd}\!=\!9.8\times 10^{-8} ({\rm \frac{m^2}{s^3}})\!=\!98 ({\rm \frac{{[\alpha m]^2}}{{[\beta s]^{3}}}})$ for case B. These values $D_{\rm fd}\!=\!9.8 ({\rm \frac{{[\alpha m]^2}}{{[\beta s]^{3}}}})$  and  $D_{\rm fd}\!=\!98 ({\rm \frac{{[\alpha m]^2}}{{[\beta s]^{3}}}})$ are comparable to the values $D\!=\!10({\rm \frac{{[\alpha m]^2}}{{[\beta s]^{3}}}})$ and $D\!=\!100 ({\rm \frac{{[\alpha m]^2}}{{[\beta s]^{3}}}})$ used in the simulations (Table \ref{table-2}).

% App C
%%%%%%%%%%%%%
\section{Discretization of kinetic energy \label{App-C}}
%%%%%%%%%%%%%
We assume that biological materials move with the flow velocity $\vec{V}$ and that their mass density is the same as that of water. Therefore, the kinetic energy of the material is given by
\begin{eqnarray}
	\label{cont-kinetic-energy}
H_K=\frac{m_0}{2}\int \sqrt{g}d^2x\; \vec{V}\cdot \vec{V}, \quad m_0=1,
	\end{eqnarray}
where $m_0\!=\!1$ is assumed in this Appendix, for simplicity. 
The discrete expression is obtained by replacing $\int \sqrt{g}d^2\vec{V}\cdot \vec{V}$ with $\sum_i (\chi_{i1}\chi_{i2})^{-1}\vec{V}_i\cdot \vec{V}_i$ for the local coordinate $(i12)$ (Fig. \ref{fig-A-1}(a)). Summing the contributions from the four possible local coordinate systems at $i$ shown in Fig. \ref{fig-A-1}(b), we obtain the factor $(\chi_{i1}\chi_{i2})^{-1}\!+\!(\chi_{i2}\chi_{i3})^{-1}\!+\!(\chi_{i3}\chi_{i4})^{-1}\!+\!(\chi_{i4}\chi_{i1})^{-1}$. This factor can be written as $4(\chi_{i1}\chi_{i2})^{-1}$ using the relations $\chi_{i1}\!=\!\chi_{i3}$ and $\chi_{i2}\!=\!\chi_{i4}$, which imply that $H_K$ is independent of the choice of local coordinates. The numerical factor $4$ can be absorbed into the normalization factor $\overline{m}$, and we obtain
\begin{eqnarray}
\label{disc-kinetic-energy}
	\begin{split}
		H_K&=\frac{1}{2}\sum_i m_{i}(\tau)\vec{V}_i\cdot \vec{V}_i, \\
		m_{i}(\tau)&=\overline {m}^{-1}\frac{1}{\chi_{i1}\chi_{i2}}, \quad\overline {m}=\left< \frac{1}{\chi_{i1}(\tau^{\rm iso})\chi_{i2}(\tau^{\rm iso})}\right>\;(\simeq  19.51\; {\rm for}\; \chi_0^m=0.01).
	\end{split}
\end{eqnarray}
The Finsler lengths $\chi_{i1}$ and $\chi_{i2}$ for $H_K$ are defined by
\begin{eqnarray}
	\label{discrete-chi-for-HK}
	\begin{split}
		\chi_{i1}&=\left\{ \begin{array}{@{\,}ll}
			\quad|\vec{\tau}_i\cdot\vec{e}_{i1}|^2+\chi_0^m\;\;\;=\cos^2\theta_{i1}+\chi_0^m\quad  ({\rm C\; model})  \\
			1-(\vec{\tau}_i\cdot\vec{e}_{i1})^2+\chi_0^m=\sin^2\theta_{i1}+\chi_0^m\quad  ({\rm S\;model})
		\end{array} \right.,\\
		\chi_{i2}&=\left\{ \begin{array}{@{\,}ll}
			\quad|\vec{\tau}_i\cdot\vec{e}_{i2}|^2+\chi_0^m\;\;\;=\cos^2\theta_{i2}+\chi_0^m\quad  ({\rm C\; model}) \\
			1-(\vec{\tau}_i\cdot\vec{e}_{i2})^2+\chi_0^m=\sin^2\theta_{i2}+\chi_0^m\quad  ({\rm S\;model})
		\end{array} \right., \\ \chi_0^m&=0.01,
	\end{split}
\end{eqnarray}
which are different from those in Eq. (\ref{discrete-chi-for-HV}) for $H_V$ due to the difference between  $\chi_0^m(=\!0.01)$ and $\chi_0^v(=\!0.5)$.

The procedure for calculating the mean value $\overline{m}$ in Eq. (\ref{disc-kinetic-energy}) is the same as that described in the paragraph below Eq. (\ref{normalization-factor}). The expression for $m_{i}(\tau)$ is common to both the S and C models and can also be expressed in terms of the angle $\theta$ between $\vec{\tau}_i$ and the $x$ axis as
\begin{eqnarray}
	\label{relative-mass-theta}
	\begin{split}
		m(\theta)&=\overline {m}^{-1}\frac{1}{(\cos^2\theta+\chi_0^m)(\sin^2\theta+\chi_0^m)},\\ \overline {m}&=\frac{1}{\pi}\int_0^\pi \frac{1}{(\cos^2\theta+\chi_0^m)(\sin^2\theta+\chi_0^m)}d\theta\;(\simeq  19.51\; {\rm for}\; \chi_0^m=0.01).
	\end{split}
\end{eqnarray}
The index $i$ is omitted because the expression with respect to  $\theta$ is the same at all lattice sites $i$.   Note that $H_K$ is not invariant under the transformation between $g_{ij}^\#$  and $g_{ij}$ in Eq. (\ref{F-metric-2D-conf})   in contrast to $H_V$ in Eq. (\ref{discrete-HV}). 
For $g_{ij}^\#$ defined in Eq. (\ref{F-metric-2D-conf}), we have another expression for $m_i$ such that
 \begin{eqnarray}
 	\label{factor-another-gauge}
 	\begin{split}
 		m^\#_{i}(\tau)&=\overline {m}^{\#-1}{\chi_{i1}\chi_{i2}},\quad \overline {m}^\#=\left< {\chi_{i1}(\tau^{\rm iso})\chi_{i2}(\tau^{\rm iso})}\right>\;(\simeq 0.1351\; {\rm for}\; \chi_0^m=0.01).
 	\end{split}
 \end{eqnarray}
 The expression using $\theta$ is given by
 \begin{eqnarray}
	\label{factor-another-gauge-theta}
	\begin{split}
		m^\#(\theta)&=\overline {m}^{\#-1}{(\cos^2\theta+\chi_0^m)(\sin^2\theta+\chi_0^m)},\\ \overline {m}^\#&=\frac{1}{\pi}\int_0^\pi {(\cos^2\theta+\chi_0^m)(\sin^2\theta+\chi_0^m)}d\theta\;(\simeq 0.1351\; {\rm for}\; \chi_0^m=0.01).
	\end{split}
\end{eqnarray}

\begin{figure}[h]
	\centering{}\includegraphics[width=9.0cm,clip]{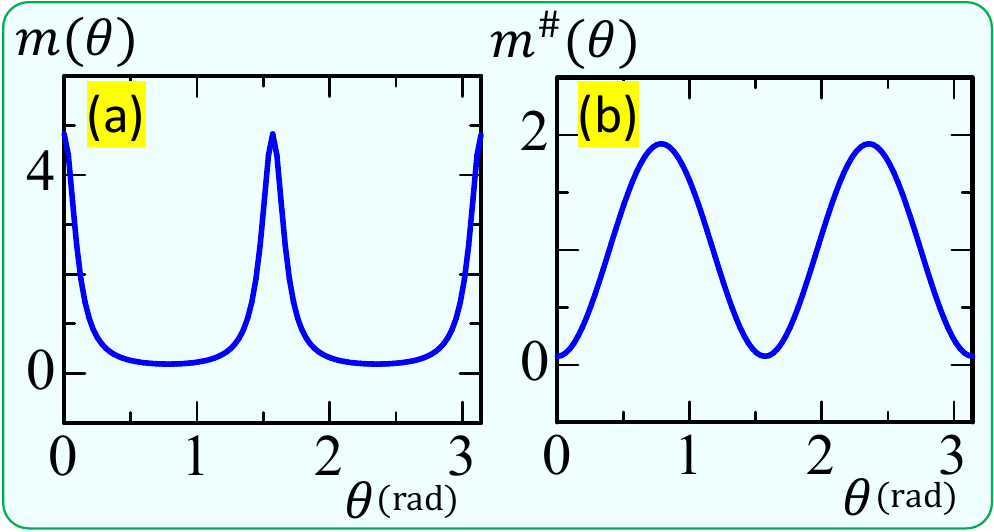}
	\caption{
		The $\theta$ dependence of relative masses (a) $m(\theta)$  and (b) $m^\#(\theta)$. $\chi_0^m$ is fixed at $\chi_0^m\!=\!0.01$.
		\label{fig-A-3}}
\end{figure}

We find from Figs. \ref{fig-A-3}(a),(b) that $m(\theta)$ has the minimum at $\theta\!\simeq\!\pi/4$ and $\theta\!\simeq\!3\pi/4$, while   $m^\#(\theta)$ has the minimum at  $\theta\!\simeq\!0,\pi$ and $\theta\!\simeq\!\pi/2$. These minimums in $m^\#(\theta)$ indicate that conformally non-flat configurations, as illustrated in Fig. \ref{fig-8}(c), appear under the gauge $g_{ij}^\#$ in Eq. (\ref{new-F-metric-2D}). 
%%%%%%%%%%%%%%%% f-A-3
%%%%%%%%%%%%%%%
At these minima, ${\boldsymbol \varGamma}_x$ or ${\boldsymbol \varGamma}_y$, and/or equivalently the relative viscosity $\Gamma_{i1}$ and/or $\Gamma_{i2}$, respectively, reach its maximum, as shown in Figs. \ref{fig-A-2}(a),(b), indicating shear-thickening behavior. However, shear thickening is not suitable for protoplasmic streaming, and hence, $g_{ij}$ in Eq. (\ref{F-metric-2D}) is assumed in this paper.

% App D
%%%%%%%%%%%%%
\section{Discretization of the advection term \label{App-D}}
%%%%%%%%%%%%%
First, we note that the flow field $\vec{V}$ can be expressed by using the basis vectors of the Cartesian coordinate and the local coordinate;
 \begin{eqnarray}
	\label{two-expressions-vector-V}
	\begin{split}
		\vec{V}=V_x\vec{e}_x+V_y\vec{e}_y=V^1\vec{e}_{i1}^{\,*}+V^2\vec{e}_{i2}^{\,*}.
	\end{split}
\end{eqnarray}
Using the basis vectors $(\vec{e}_{i1}^{\,*}, \vec{e}_{i2}^{\,*})$ corresponding to the local coordinates plotted in Fig. \ref{fig-A-1}(b)), we have the Cartesian basis vectors 
 \begin{eqnarray}
	\label{relation-basis-vectors}
	\begin{split}
		\vec{e}_x&=\frac{\vec{e}_{i1}^{\,*}}{\sqrt{\vec{e}_{i1}^{\,*}\cdot \vec{e}_{i1}^{\,*}}}=\frac{\vec{e}_{i1}^{\,*}}{\sqrt{g_{11}}}, \quad 	\vec{e}_y=\frac{\vec{e}_{i2}^{\,*}}{\sqrt{\vec{e}_{i2}^{\,*}\cdot \vec{e}_{i2}^{\,*}}}=\frac{\vec{e}_{i2}^{\,*}}{\sqrt{g_{22}}}, \quad (i12), \\
		\vec{e}_x&=-\frac{\vec{e}_{i2}^{\,*}}{\sqrt{\vec{e}_{i2}^{\,*}\cdot \vec{e}_{i2}^{\,*}}}=-\frac{\vec{e}_{i2}^{\,*}}{\sqrt{g_{22}}}, \quad 	\vec{e}_y=\frac{\vec{e}_{i1}^{\,*}}{\sqrt{\vec{e}_{i1}^{\,*}\cdot \vec{e}_{i1}^{\,*}}}=\frac{\vec{e}_{i1}^{\,*}}{\sqrt{g_{11}}}, \quad (i23), \\
		\vec{e}_x&=-\frac{\vec{e}_{i1}^{\,*}}{\sqrt{\vec{e}_{i1}^{\,*}\cdot \vec{e}_{i1}^{\,*}}}=-\frac{\vec{e}_{i1}^{\,*}}{\sqrt{g_{11}}}, \quad 	\vec{e}_y=-\frac{\vec{e}_{i2}^{\,*}}{\sqrt{\vec{e}_{i2}^{\,*}\cdot \vec{e}_{i2}^{\,*}}}=-\frac{\vec{e}_{i2}^{\,*}}{\sqrt{g_{22}}}, \quad (i34), \\
		\vec{e}_x&=\frac{\vec{e}_{i2}^{\,*}}{\sqrt{\vec{e}_{i2}^{\,*}\cdot \vec{e}_{i2}^{\,*}}}=\frac{\vec{e}_{i2}^{\,*}}{\sqrt{g_{22}}}, \quad 	\vec{e}_y=-\frac{\vec{e}_{i1}^{\,*}}{\sqrt{\vec{e}_{i1}^{\,*}\cdot \vec{e}_{i1}^{\,*}}}=-\frac{\vec{e}_{i1}^{\,*}}{\sqrt{g_{11}}}, \quad (i41),
	\end{split}
\end{eqnarray}
which are calculated as the unit vectors using  $(\vec{e}_{i1}^{\,*}, \vec{e}_{i2}^{\,*})$. Combining these $(\vec{e}_x,\vec{e}_y)$ and $\vec{V}$ in Eq. (\ref{two-expressions-vector-V}), we find
 \begin{eqnarray}
	\label{relation-coefficients-vector}
	\begin{split}
		V^1&=\sqrt{g^{11}}V_x, \quad V^2=\sqrt{g^{22}}V_y, \quad (i12), \\
		V^1&=\sqrt{g^{11}}V_y, \quad V^2=-\sqrt{g^{22}}V_x, \quad (i23), \\
		V^1&=-\sqrt{g^{11}}V_x, \quad V^2=-\sqrt{g^{22}}V_y, \quad (i34), \\
		V^1&=-\sqrt{g^{11}}V_y, \quad V^2=\sqrt{g^{22}}V_x, \quad (i41),
	\end{split}
\end{eqnarray}
where $g_{ij}g^{jk}\!=\!\delta_{ik}$ are used.

The relations in Eq. (\ref{relation-coefficients-vector}) are used to obtain the dicsretization of 
the advection term
 \begin{eqnarray}
	\label{cont-convex-term}
\left(\vec{V}\cdot \nabla\right)\vec{V}.
\end{eqnarray}
For this purpose, the contravariant basis vectors $(\vec{e}^{\,i1}_{*}, \vec{e}^{\,i2}_{*})$ for the local coordinate $(i12)$ at vertex $i$ are used; 
 \begin{eqnarray}
	\label{relation-covariant-contravariant-basis}
	\begin{split}
	\vec{e}^{\,i1}_{*}\cdot \vec{e}_{i1}^{\,*}&=\vec{e}_{i1}^{\,*}\cdot \vec{e}^{\,i1}_{*}=\vec{e}^{\,i2}_{*}\cdot \vec{e}_{i2}^{\,*}=\vec{e}_{i2}^{\,*}\cdot \vec{e}^{\,i2}_{*}=1,\\
\vec{e}^{\,i1}_{*}\cdot \vec{e}_{i2}^{\,*}&=\vec{e}_{i1}^{\,*}\cdot \vec{e}^{\,i2}_{*}=\vec{e}^{\,i2}_{*}\cdot \vec{e}_{i1}^{\,*}=\vec{e}_{i2}^{\,*}\cdot \vec{e}^{\,i1}_{*}=0.
	\end{split}
\end{eqnarray}
Since $\nabla$ can also be expressed as  $\nabla\!=\!\vec{e}^{\,i1}_{*}\frac{\partial}{\partial x^1}\!+\!\vec{e}^{\,i2}_{*}\frac{\partial}{\partial x^2}$ using $(\vec{e}^{\,i1}_{*}, \vec{e}^{\,i2}_{*})$, we find a local coordinate expression of
 $\vec{V}\cdot \nabla$ at vertex $i$ such that
 \begin{eqnarray}
	\label{V-cdot-nabla}
	\begin{split}
\vec{V}\cdot \nabla&=\left(V^j  {\vec{e}_{ij}^{\,*}}\right)\cdot\left({\vec{e}^{\,ik}_{*}}\frac{\partial}{\partial x^k}\right)
=\delta_{jk}V^j\frac{\partial}{\partial x^k}=V^j\frac{\partial}{\partial x^j}\\
&=V^1\frac{\partial}{\partial x^1}+V^2\frac{\partial}{\partial x^2}.
	\end{split}
\end{eqnarray}
Multiplying this operator to $\vec{V}_i$ and replacing $\frac{\partial}{\partial x^1}\vec{V}_i$ with 
$\vec{V}_1\!-\!\vec{V}_i$ and  $\frac{\partial}{\partial x^2}\vec{V}_i$ with 
$\vec{V}_2\!-\!\vec{V}_i$, we find 
 \begin{eqnarray}
	\label{disc-convex-term-i12}
	\begin{split}
	\left(\vec{V}\cdot \nabla\right)\vec{V}&\to V^1(\vec{V}_1-\vec{V}_i)+V^2(\vec{V}_2-\vec{V}_i)\\
	&=\sqrt{g^{11}}V_x(\vec{V}_1-\vec{V}_i)+\sqrt{g^{22}}V_y(\vec{V}_2-\vec{V}_i),\quad (i12).
	\end{split}
\end{eqnarray}
The other three expressions are given by
 \begin{eqnarray}
	\label{disc-convex-term-other}
	\begin{split}
		&\sqrt{g^{11}}V_y(\vec{V}_2-\vec{V}_i)-\sqrt{g^{22}}V_x(\vec{V}_3-\vec{V}_i),\quad (i23),\\
-&\sqrt{g^{11}}V_x(\vec{V}_3-\vec{V}_i)-\sqrt{g^{22}}V_y(\vec{V}_4-\vec{V}_i),\quad (i34),\\
-&\sqrt{g^{11}}V_y(\vec{V}_4-\vec{V}_i)+\sqrt{g^{22}}V_x(\vec{V}_1-\vec{V}_i),\quad (i41).
	\end{split}
\end{eqnarray}
By including these terms in $\left(\vec{V}\cdot \nabla\right)\vec{V}$ with the numerical factor $1/4$, $\left(\vec{V}\cdot \nabla\right)\vec{V}$ can be written as
 \begin{eqnarray}
	\label{disc-convex-term-other-1}
	\begin{split}
		\left(\vec{V}\cdot \nabla\right)\vec{V}\to&\frac{1}{4}\left(\sqrt{g^{22}}(i41)+\sqrt{g^{11}}(i12)\right)V_x(\vec{V}_1-\vec{V}_i) \\
		+&\frac{1}{4}\left(\sqrt{g^{22}}(i12)+\sqrt{g^{11}}(i23)\right)V_y(\vec{V}_2-\vec{V}_i)\\
		-&\frac{1}{4}\left(\sqrt{g^{22}}(i23)+\sqrt{g^{11}}(i34)\right)V_x(\vec{V}_3-\vec{V}_i)\\
		-&\frac{1}{4}\left(\sqrt{g^{22}}(i34)+\sqrt{g^{11}}(i41)\right)V_y(\vec{V}_4-\vec{V}_i).
	\end{split}
\end{eqnarray}
Rewriting the coefficients explicitly, we have
 \begin{eqnarray}
	\label{disc-convex-term-other-2}
	\begin{split}
		\left(\vec{V}\cdot \nabla\right)\vec{V}\to&\frac{1}{4}\left(\chi_{i1}+\chi_{i1}\right)V_x(\vec{V}_1-\vec{V}_i) 
		+\frac{1}{4}\left(\chi_{i2}+\chi_{i2}\right)V_y(\vec{V}_2-\vec{V}_i)\\
		-&\frac{1}{4}\left(\chi_{i3}+\chi_{i3}\right)V_x(\vec{V}_3-\vec{V}_i)
		-\frac{1}{4}\left(\chi_{i4}+\chi_{i4}\right)V_y(\vec{V}_4-\vec{V}_i).
	\end{split}
\end{eqnarray}
Since $\chi_{i1}\!=\!\chi_{i3}$ and $\chi_{i2}\!=\!\chi_{i4}$, we obtain 
\begin{eqnarray}
	\label{disc-convex-term-other-3}
%	\begin{split}
		\left(\vec{V}\cdot \nabla\right)\vec{V}\to\frac{1}{2}\chi_{i1}V_x(\vec{V}_1-\vec{V}_3)+\frac{1}{2}\chi_{i2}V_y(\vec{V}_2-\vec{V}_4).
%			\end{split}
\end{eqnarray}
Thus, by introducing a normalization factor $\overline{C}$, we obtain for $\vec{V}_i$ at vertex $i$ that
\begin{eqnarray}
	\label{disc-convex-term}
	\begin{split}
		\left(\vec{V}_i\cdot \nabla\right)\vec{V}_i&=\frac{1}{2}C_{i1}V_{i,x}(\vec{V}_1-\vec{V}_3)+\frac{1}{2}C_{i2}V_{i,y}(\vec{V}_2-\vec{V}_4),\\
		C_{ij}&=\overline{C}^{-1}\chi_{ij}, (j=1,2), \quad \overline{C}=\langle \chi_{i1}(\tau^{\rm iso})\rangle=\langle \chi_{i2}(\tau^{\rm iso})\rangle.
	\end{split}
\end{eqnarray}
For isotropic $\tau^{\rm iso}$, it is expected that $C_{ij}(\tau^{\rm iso})\!\to\!1$. Therefore, we have
\begin{eqnarray}
	\label{disc-convex-term-cartesian}
		\left(\vec{V}_i\cdot \nabla(\tau^{\rm iso})\right)\vec{V}_i=\frac{1}{2} V_x(\vec{V}_1-\vec{V}_3)+\frac{1}{2}V_y(\vec{V}_2-\vec{V}_4).
\end{eqnarray}
This expression is identical with the case that $\vec{V}_i\cdot \nabla$ is described by the Cartesian coordinate expression.


\begin{thebibliography}{9}


% ------- cytoplasmic streaming
\bibitem{ShimmenYokota-COCB2004}
T. Shimmen and T. Yokota, {\it Cytoplasmic streaming in plants}, Current Opinion in Cell Biology, {\bf 16 (1)}, 68--72 (2004); https://doi.org/10.1016/j.ceb.2003.11.009.

\bibitem{VLubics-Goldstein-Protoplasma2009}
J. Verchot-Lubicz, and R.E.Goldstein, 
{\it Cytoplasmic streaming enables the distribution of molecules and vesicles in large plant cells}, 
Protoplasma, {\bf 240}, 99-107.(2009); DOI 10.1007/s00709-009-0088-x.

\bibitem{TominagaIto-COPB2015}
M. Tominaga and K. Ito, {\it The molecular mechanism and physiological role of cytoplasmic streaming}, Current Opinion in Plant Biology, {\bf 27}, 104--110 (2015); https://doi.org/10.1016/j.pbi.2015.06.017.


% ------- protoplasmic streaming
\bibitem{Kamiya-Kuroda-1956}
N. Kamiya and K. Kuroda, {\it Velocity Distribution of the Protoplasmic Streaming in Nitella Cells}, 
Bot, Mag. Tokyo, {\bf 69}, 544-554 (1956); https://doi.org/10.15281/jplantres1887.69.544.

\bibitem{Kamiya-Kuroda-1958}
N. Kamiya and K. Kuroda, {\it Measurement of the Motive Force of the Protoplasmic Rotation in Nitella}, 
Protoplasma, {\bf 50}, 144-147 (1958). 

\bibitem{Kamiya-Kuroda-1973}
N. Kamiya and K. Kuroda, {\it Dynamics of Cytoplasmic Streaming in a Plant Cell}, 
Biorheology, {\bf 10}, 179-187 (1973). 

\bibitem{Kamiya-1986}
N. Kamiya, {\it Cytoplasmic streaming in giant algal cells: A historical survey of experimental approaches},
Bot, Mag. Tokyo, {\bf 99}, 441-496 (1986); 
https://doi.org/10.1007/BF02488723. 

\bibitem{Tazawa-pp1968}
M. Tazawa, {\it Motive force of the cytoplasmic streaming in Nitella}, Protoplasma, {\bf 65}, 207-222 (1968).


% ------- physics experimental velocity distribution
\bibitem{Mustacich-Ware-PRL1974}
R.V. Mustacich and B.R. Ware, {\it Observation of Protoplasmic Streaming by Laser-Light Scattering}, 
Phys. Rev. Lett. {\bf 33}, 617-620 (1974). 

\bibitem{Mustacich-Ware-BJ1976}
R.V. Mustacich and B.R. Ware, {\it A Study of Protoplasmic Streaming in Nitella by Laser Doppler spectroscopy}, 
Boiophys. J. {\bf 16}, 373-388 (1976). 

\bibitem{Mustacich-Ware-BJ1977}
R.V. Mustacich and B.R. Ware, {\it Velocity Distributions of the Streaming Protoplasm in Nittella Flexilis}, 
Boiophys. J. {\bf 17}, 229-241 (1977). 

\bibitem{Sattelle-Buchan-JCS1976}
D.B. Sattelle and P.B. Buchan, {\it Cytoplasmic Streaming in Chara Corallina studied by Laser Light Scattering}, 
J. Cell. Sci. {\bf 22}, 633-643 (1976). 


% ------ molecular motor
\bibitem{Squires-Quake-RMP2005}
T.M. Squires and S.R. Quake,
{\it Microfluidics: Fluid physics at the nanoliter scale.},
Rev. Mod. Phys.{\bf 77}, 977--1026 (2005);
https://doi.org/10.1103/RevModPhys.77.977.

\bibitem{McintoshOstap-CSatAG2016}
B.B. McIntosh and E.M. Ostap, 
{\it Myosin-I molecular motors at a glance}, 
Cell Science at a Glance, {\bf 129}, 2689--2695 (2016); 
https://doi:10.1242/jcs.186403.

\bibitem{Astumian-Science2020}
R.D. Astumian,
{\it Thermodynamics and Kinetics of a Brownian Motor},
Science, {\bf 276}, 917--922 (2020); 
http://science.sciencemag.org/content/276/5314/917.

\bibitem{Julicher-etal-RMP1997}
F. J${\ddot{\rm u}}$licher, A. Ajdari and J. Prost,
{\it Modeling molecular motors},
Rev. Mod. Phys. {\bf 69(4)}, 1269--1282 (1997);
https://link.aps.org/doi/10.1103/RevModPhys.69.1269.


% ----- Brownian Dynamics
\bibitem{Ermark-McCammon-JCP1978}
D.L. Ermark and J.A. McCammon, 
{\it Brownian dynamics with hydrodynamic interactions},
J. Chem. Phys. {\bf 69}, 1352 (1978); 
https://doi.org/10.1063/1.436761.

\bibitem{Brady-McCammon-ARFM1988}
J. F. Brady and G. Bossis, 
{\it Stokesian Dynamics},
Ann. Rev. Fluid Mech. {\bf 20}, pp. 111-57 ( 1988); 
https://doi.org/10.1146/annurev.fl.20.010188.000551.

\bibitem{Roux-PhysA1992}
J. N. Roux, {\it Brownian particles at different times scales: a new derivation of the Smoluchowski equation},
Physica A {\bf 188}, pp. 526-552, (1992); 
https://doi.org/10.1016/0378-4371(92)90330-S.

% ------- Langevin sim. Brownian dynamics
\bibitem{Hossain-etal-POF2022}
M.T. Hossain, I.D. Gates and G. Natale, {\it Dynamics of Brownian Janus rods at a liquid-liquid interface}, Phys. Fluids 34, 012117 (2022); https://doi.org/10.1063/5.0076148.

% ------ physical properties 
\bibitem{Amari-etal-BioPhysi2025}
A. Toshiki, N. Nagata, M. Tominaga, H. Takatsuka,
{\it Cytoskeleton as a generator of characteristic physical properties of plant cells: 'cell wall,' 'large vacuole,' and 'cytoplasmic streaming'},
Biophysics and Physicobiology, {\bf 22}, e220013 (2025);
https://doi.org/10.2142/biophysico.bppb-v22.0013.

% ------- streaming experimentally
\bibitem{Goldstein-etal-JFM2010}
J-W. van De Meent, A.J. Sederman, L.F. Gladden and R.E. Goldstein, {\it Measurement of cytoplasmic streaming in single plant cells by magnetic resonance velocimetry}, J. Fluid Mech. {\bf 642}, pp.5-14 (2010); doi:10.1017/S0022112009992187

\bibitem{Raymond-Goldstein-IF2015}
R.E. Goldstein and J-W. van de Meent, {\it Physical perspective on cytoplasmic streaming}, 
Interface Focus. {\bf 5}: 20150030; 
https://doi.org/10.1098/rsfs.2015.0030. 

% ------- streaming experimentally
\bibitem{Kikuchi-Mochizuki-PlosOne2015}
K. Kikuchi and O. Mochizuki, {\it Diffusive Promotion by Velocity Gradient of Cytoplasmic Streaming (CPS) in Nitella Internodal Cells}, Plos One, 0144938(1-12) (2015); 
https://DOI:10.1371/journal.pone.0144938. 

% ------- streaming numerically
\bibitem{Niwayama-etal-PNAS2010}
R. Niwayama, K. Shinohara and A. Kimura, {\it Hydrodynamic property of the cytoplasm is sufficient to mediate cytoplasmic streaming in the Caenorhabiditis elegans embryo}, 
PNAS, vol. {\bf 108}, pp.11900-11905 (2011);
https://doi.org/10.1073/pnas.1101853108.

% ------- microtubule 
\bibitem{Ganguly-Goldstein-etal-PNAS2012}
S. Gangulya, L.S. Williams, I.M. Palacios, and R.E. Goldstein, {\it Cytoplasmic streaming in Drosophila oocytes varies with kinesin activity and correlates with the microtubule cytoskeleton architecture}, PNAS vol {\bf 109}, pp.15109–15114 (2012); https://www.pnas.org/cgi/doi/10.1073/pnas.1203575109.


% ------ physical properties role of wall slip 
\bibitem{Wolff-etal-Interface2012}
K. Wolff, D. Marenduzzo and M.E. Cates, {\it Cytoplasmic streaming in plant cells: the role  of wall slip},
J. R. Soc. Interface {\bf 9}, 1398-1408 (2012); https://doi.org/10.1098/rsif.2011.0868.

% ------ cytoplasmic streaming elongated cells 
\bibitem{Htet-Lauga-PNASNexus2025}
P.H. Htet and E. Lauga, {\it Analytical methods for cytoplasmic streaming in elongated cells},
PNAS Nexus {\bf 4}, pgaf057 (2025); https://doi.org/10.1093/pnasnexus/pgaf057.

% ------- streaming theoretical
\bibitem{Goldstein-etal-PRL2008}
J-W. Meent, I. Tuvalk and R.E. Goldstein, {\it Nature's Microfluidic Transporter: Rotational Cytoplasmic Streaming at High P$\acute {\rm e}$clet Numbers}, Phys. Rev. Lett. {\bf 101}, 178102(1-4) (2008); https://doi.org/10.1103/PhysRevLett.101.178102.

\bibitem{Goldstein-etal-PNAS2008}
R.E. Goldstein, I. Tuvalk and J-W. van de Meent, {\it Microfluidics of cytoplasmic streaming and its implications for intracellular transport},
PNAS, {\bf 105}, 3663-3667 (2008); 
https://www.pnas.org/cgi/doi/10.1073 pnas.0707223105.


% polymer physics
\bibitem{doi2013soft}
M. Doi, Soft matter physics (Oxford University Press, 2013), https://doi.org/10.1080/1358314X.2014.912382.
%
\bibitem{Doi-Edwards-1986}
M. Doi and S.F. Edwards, {\it The Theory of Polymer Dynamics}; Oxford University Press: Oxford, UK, 1986.


% ----- polymer-solution-rheology
\bibitem{Schroeder-JRheo2018}
C.M. Schroeder, {\it Single polymer dynamics for molecular rheology}, Journal of Rheology {\bf 62}, 371 (2018); https://doi.org/10.1122/1.5013246

\bibitem{Chabra-Richardson-2025}
R.P. Chhabra and J.F. Richardson, {\it Non-Newtonian Flow and Applied Rheology: Engineering Applications}, Second edition,  (Butterworth-Heinemann, 2008); https://doi.org/10.1016/B978-0-7506-8532-0.X0001-7.

\bibitem{Dinic-etal-Wiley2022}
J. Dinic, C.D.V.M. Narváez and V. Sharma, {\it Rheology of Unentangled Polymer Solutions Depends on  Three Macromolecular Properties: Flexibility, Extensibility, and Segmental Dissymmetry}, In {\it  Macromolecular Engineering: From Precise Synthesis to Macroscopic Materials and Applications}, Second Edition, Edited by K. Matyjaszewski, Y. Gnanou, N. Hadjichristidis and M. Muthukumar, (Wiley, 2022), DOI: 10.1002/9783527815562.mme0067

% ------- Langevin NS for the streaming
\bibitem{Egorov-etal-POF2020}
V. Egorov. O. Maksimova, I. Andreeva, H. Koibuchi, S. Hongo, S. Nagahiro, H. Ikai, M. Nakayama, S. Noro, T. Uchimoto and J-P. Rieu, {\it Stochastic fluid dynamics simulations of the velocity distribution in protoplasmic streaming}, Phys. Fluids, {\bf 32}, 121902(1-15)  (2020); 
https://doi.org/10.1063/5.0019225.

\bibitem{Noro-etal-PHYSA2023}
S. Noro, S. Hongo, S. Nagahiro, H. Ikai, H. Koibuchi, M. Nakayama, T. Uchimoto and  G. Diguet, {\it Langevin and Navier–Stokes simulation of three-dimensional protoplasmic streaming},
Physica A {\bf 628}  129154  (2023); https://doi.org/10.1016/j.physa.2023.129154.

% Finsler geometry modeling

\bibitem{Koibuchi-Sekino-PhysA2014} 
H. Koibuchi and H. Sekino, {\it Monte Carlo studies of a Finsler geometric surface model}, {Physica A},   {\bf 303}, 37-50 (2014); https://doi.org/10.1016/j.physa.2013.08.006.

\bibitem{Proutorov-etal-JPC2018} 
E. Proutorov, N. Matsuyama and H. Koibuchi, {Finsler geometry modeling and Monte Carlo study of liquid crystal elastomers under electric fields}, {\it J. Phys. Cond. Mat.},  {\bf 30} (2018) 405101, https://doi.org/10.1088/1361-648X/aadcba.

\bibitem{FKato-etal-PhysA2026} 
F. Kato, H. Koibuchi, M. Nakayama, S. Tasaki and T. Uchimoto, {\it Turing patterns on non-fluctuating surfaces under mechanical
	stresses}, {Physica A}, {\bf 703}  132137  (2026); https://doi.org/10.1016/j.physa.2026.132137.

\bibitem{SS-Chern-AMS1996}
S.-S. Chern, {Finsler Geometry Is Just Riemannian Geometry without the Quadratic Restriction}, In \textit{Notices of the AMS},  959-963  (1996);    https://www.ams.org/notices/199609/chern.pdf.


% ------ radial angle dependence
\bibitem{Pickard-CJB1971}
W. F. Pickard, {\it Further observations on cytoplasmic streaming in Chara braunii}, Can. J. Bot. {\bf 50}, 703-711 (1971).










% ------ note-1
\bibitem{note-1}
Flows between parallel plates are considered two-dimensional if there is no flow in the third direction, perpendicular to the two-dimensional domain (Fig. \ref{fig-2}(a)). However, this picture is no longer valid for fluids containing polymeric materials, which cannot rotate solely within the two-dimensional domain.












% ------- MAC
\bibitem{McKee-etal-CandF2007}
S. McKee, M.F. Tom$\acute{\rm e}$, V.G. Ferreira, J.A. Cuminato, A. Castelo, F.S. Sousa, N. Mangiavacchi, \newblock{{T}he {MAC} method}. \newblock Computers $\&$ Fluids 2008;37: 907-930;
doi:10.1016/j.compfluid.2007.10.006.


% -----Landau-Lifschitz
\bibitem{Landau-Lifschitz-StatPhys}
L.D. Landau and E.M. Lifschitz,
{\it Course of Theoretical Phisics}, Vol. 9, Statistical Physics, Part 2,
Chapter IX, Hydrodynamic Fluctuations.

%---- Dissipative Particle Dynamics
\bibitem{Hoogerbrugge-Koleman-EPL1992}
P.J. Hoogerbrugge and J.M. Koleman, 
{\it Simulating Microscopic Hydrodynamic Phenomena with Dissipative Particle Dynamics},
Europhys. Lett., {\bf 19} (3), pp. 155-160 (1992); 
DOI 10.1209/0295-5075/19/3/001.

\bibitem{Espanol-Warren-EPL1995}
P. Espa${\rm \tilde{n}}$ol and P. Warren, 
{\it Statistical Mechanics of Dissipative Particle Dynamics},
Europhys. Lett., {\bf 30} (4), pp. 191-196 (1995); 
DOI 10.1209/0295-5075/30/4/001.

\bibitem{Espanol-EPL1997}
P. Espa${\rm \tilde{n}}$ol, 
{\it Dissipative particle dynamics with energy conservation},
Europhys. Lett., {\bf 40} (6), pp. 631-636 (1997); 
DOI 10.1209/epl/i1997-00515-8.


\bibitem{Groot-Warren-JCP1997}
R.D. Groot and P.B. Warren, 
{\it Dissipative particle dynamics: Bridging the gap between atomistic and mesoscopic simulation},
J. Chem. Phys. {\bf 107}, 4423 (1997); https://doi.org/10.1063/1.474784


\bibitem{Hoogerbrugge-Koelman-EPL1992}
P.J. Hoogerbrugge and  J.M.V.A. Koelman, 
{\it Simulating Microscopic Hydrodynamic Phenomena with Dissipative Particle Dynamics},
Europhys. Lett., {\bf 19} (3), pp. 155-160 (1992); 
DOI 10.1209/0295-5075/19/3/001.

\bibitem{Nikunen-etal-CPC2003}
P. Nikunen, M. Karttunen and I. Vattulainen, 
{\it How would you integrate the equations of motion in dissipative particle dynamics simulations?},
Computer Physics Communications {\bf 153} 407–423  (2003); 
doi:10.1016/S0010-4655(03)00202-9.









% ------ MC technique
\bibitem{Metropolis-JCP-1953} 
N. Metropolis, A.W. Rosenbluth, M.N. Rosenbluth, and A.H. Teller, 
{\it Equation of State Calculations by Fast Computing Machines},
J. Chem. Phys. \textbf{21}, 1087 (1953);
https://doi.org/10.1063/1.1699114.

\bibitem{Landau-PRB1976} 
D.P. Landau, {\it Finite-size behavior of the simple-cubic Ising lattice},
Phys. Rev. B \textbf{13}, 2997 (1976); DOI:https://doi.org/10.1103/PhysRevB.14.255.





% ------- conformal
\bibitem{Andersen-etal-NatPhys2025}
B.H. Andersen, F.M.R. Safara, V. Grudtsyna,
O.J. Meacock, S.G. Andersen, W.M. Durham,
N.A.. Araujo and A. Doostmohammadi,  {\it Evidence of universal conformal invariance
	in living biological matter}, Nature Physics, vol {\bf 21} 618-623 (2025);  https://doi.org/10.1038/s41567-025-02791-2.














\end{thebibliography}
\end{document}